\documentclass[aps,prd,showpacs,amssymb,amsmath,nofootinbib]{revtex4}
\usepackage{graphicx}

\begin{document}
\title{Probing Quantized Einstein-Rosen Waves with Massless Scalar Matter}

\author{J. Fernando \surname{Barbero G.}}
\email[]{fbarbero@iem.cfmac.csic.es} \affiliation{Instituto de
Estructura de la Materia, CSIC, Serrano 123, 28006 Madrid, Spain}
\author{I\~naki  \surname{Garay}}
\email[]{igael@iem.cfmac.csic.es} \affiliation{Instituto de
Estructura de la Materia, CSIC, Serrano 123, 28006 Madrid, Spain}
\author{Eduardo J. \surname{S. Villase\~nor}}
\email[]{ejsanche@math.uc3m.es} \affiliation{Grupo de
Modelizaci\'on y Simulaci\'on Num\'erica, Universidad Carlos III
de Madrid, Avda. de la Universidad 30, 28911 Legan\'es, Spain}
\affiliation{Instituto de Estructura de la Materia, CSIC, Serrano
123, 28006 Madrid, Spain}

\date{April 19, 2006}

\begin{abstract}
The purpose of this paper is to discuss in detail the use of
scalar matter coupled to linearly polarized Einstein-Rosen waves
as a probe to study quantum gravity in the restricted setting
provided by this symmetry reduction of general relativity. We will
obtain the relevant Hamiltonian and quantize it with the
techniques already used for the purely gravitational case. Finally
we will discuss the use of particle-like modes of the quantized
fields to operationally explore some of the features of quantum
gravity within this framework. Specifically we will study
two-point functions, the Newton-Wigner propagator, and radial wave
functions for one-particle states.
\end{abstract}


\maketitle

\section{Introduction}{\label{Intro}}

There are reasons to believe that general relativity is not the
last word as far as gravitational physics is concerned, for one
thing, we still do not know how to reconcile it with quantum
mechanics, the other fundamental pillar of our description of
physical reality. As of today we lack a complete and consistent
description of quantum gravity although some proposals have made
steady progress towards our understanding of the problem if not of
its solution. The questions to be addressed by quantum gravity
must go beyond the mere finding of a consistent quantization of
general relativity because it is possible that many of the
physical concepts that work classically are inappropriate in the
quantum setting. Let us consider, for example, the metric. Though
this is the fundamental concept in classical general relativity,
formalisms such as loop quantum gravity suggest that it may not be
the basic variable at the quantum level. It is perfectly
conceivable that the metric is some kind of semiclassical
construct that emerges at scales much larger than the Planck
scale. We are left, then, with the problem of understanding what
the relevant object at the shortest scales is and how a metric
description appears. Another concept that may not be appropriate
is that of the test particle. Its classical meaning is clear, they
are point-like objects of negligible mass and, hence, with no
influence as sources for the gravitational field described by a
metric whose geodesics are the trajectories followed by these test
particles in their motion through space and time. They are, in a
definite sense, the \textit{tools} that we have available to
extract the non-trivial geometric content from the metric and
describe spacetime physics. In the context of general relativity
it is very useful and illuminating to adopt an operational point
of view to define geometric quantities such as time intervals,
lengths, and other geometrical objects with an immediate physical
interpretation. In this way it is possible to avoid resting too
much on our classical (non-relativistic) intuition and isolate the
basic and relevant physical concepts. If a consistent quantization
of gravity was available it would seem necessary to replace test
particles by \textit{quantum} objects. One of the obvious choices
--though possibly not the only one-- is to introduce quantum
particles moving in the non-trivial background defined by the
metric. The fact that this will in general be curved would
manifest itself in the way wave functions spread throughout
spacetime. Another possibility would be to couple \textit{test
fields} and use their quanta as test particles, yet many others
may exist as we cannot be sure that the correct description at
quantum gravity scales will be that provided by particle-like
objects.

The purpose of this paper is to explore these issues in the
restricted setting provided by the linearly polarized two-Killing
vector reductions of general relativity, the so called
Einstein-Rosen waves. The general problem of introducing quantum
test fields consistently is a non-trivial one. We will see,
however, that it is possible to use some fields for this purpose
because we will be able to solve the theory exactly both at the
classical and quantum level even in their presence. Specifically
we will expand here the results presented in
\cite{BarberoG.:2005ge} showing that Einstein-Rosen waves coupled
to a massless scalar field can be exactly quantized. After this we
will use the particle-like quanta of this scalar field as a way to
extract information about this quantum gravity toy model in an
operational way, much in the same way as test particles are used
to derive spacetime physics in general relativity. Our concern
will be the recovery of classical trajectories for particles and
an approximate metric description. The results obtained here
complement in a sense those already derived by considering the
microcausality of this model \cite{BarberoG.:2003ye,
BarberoG:2004uc, BarberoG.:2004uv} that clearly show how a long
distance limit appears in which microcausality reduces to the
familiar one defined by the Minkowski metric.

\bigskip

The structure of the paper is the following. After this
introduction we describe in section \ref{Eplusmatter} the
classical solution of a system of linearly polarized
Einstein-Rosen waves coupled to a massless, cylindrically
symmetric, scalar field. We will use the Geroch formalism
\cite{Geroch:1970nt} and take advantage of the two Killing vector
fields available in order to describe the model as a 1+1
dimensional system of coupled fields. As we will see the structure
of the solutions to the field equations strongly suggests that the
Hamiltonian of the system has a very simple form. We discuss this
in section \ref{Hamil}.  One of the points that merits special
attention here is the treatment of the boundary terms that give
rise to the Hamiltonian. In fact, the most salient feature of the
model is the appearance of a Hamiltonian that is a non-trivial
bounded function of the sum of the free Hamiltonians corresponding
to two free, massless, axially symmetric scalars in a 2+1
dimensional Minkowskian background. This fact will eventually
allow us to quantize the model even though it is non-trivial and
interacting. In section \ref{quant} we apply well-known Fock space
techniques, similar to the ones already used in
\cite{Ashtekar:1996bb}, to study its quantization. After this we
will describe several applications. We start in section \ref{two}
by studying the two-point functions. They can be interpreted as
approximate probability amplitudes for a particle created at a
certain radial distance from the cylindrical symmetry axis at a
certain time to be detected somewhere else at a different instant
of time. The comparison of these amplitudes with the ones
corresponding to a cylindrically symmetric massless scalar in a
Minkowski background shows some characteristic features due to the
presence of gravity such as an enhancement of the probability of
finding the field quanta close to the axis. Microcausality is
easily reviewed in this framework because the field commutator can
be obtained from these two-point functions. As the reduced field
model has two different fields, the ``gravitational" and the
``matter" scalars, we can study their commutators. We will see
that they are not zero as a consequence of the non-trivial
interacting character of the model; the commutator does not show
the microcausal behavior that appears when only one type of field
is considered.

The main reason why the interpretation of the above two-point
functions as probability amplitudes is only approximate is the
fact that the states obtained by acting with the field operators
on the vacuum are not orthonormal. The obtention of an orthonormal
basis of position states in quantum field theory is a difficult
problem that was essentially solved by Newton and Wigner for some
special types of fields \cite{NW}. By using this basis it is
possible to introduce relativistic wave functions whose modulus
squared can be interpreted as probabilities to find a particle
around a certain spacetime point.\footnote{Although these so
called Newton-Wigner states provide a rather complete solution to
the problem of localization in some relativistic quantum field
theories they also display some disturbing features mainly related
to their behavior under Lorentz transformations
\cite{Teller:1995ti}.} We introduce analogous objects in section
\ref{NWig}. Due to the absence of Lorentz invariance in the
present model some of the unpleasant features of the ordinary
Newton-Wigner states are absent here. We use these states in a
double fashion. First we study the propagator $\langle
R_2|\hat{U}(t_2,t_1)|R_1\rangle$, that can be interpreted now as a
proper probability amplitude, and compare it with the two-point
function considered above. After this we define ``radial wave
functions" in terms of these Newton-Wigner states in section
\ref{Wave} and study in detail a certain family of them for which
the evolution can be explicitly obtained in closed form as a
one-dimensional integral. This provides us with the main tool that
we have been searching for in the paper: a way to introduce
quantized particles with controlable wave functions that we can
use as quantum test particles to explore the spacetime physics of
this quantum gravitational toy model. We will use them to study
the propagation of particles in this purely quantum spacetime.

Throughout the last sections of the paper we will make use of
asymptotic techniques developed by the authors in
\cite{BarberoG:2004uc} to extract relevant information from the
closed integral expressions that describe two-point functions,
propagators, and wave functions. This is useful for two different
reasons. In some cases it allows us to consider the ``macroscopic"
(or ``classical") limit in which we only consider the behavior of
the relevant objects at scales much larger than the quantum
gravitational one. In others they provide us with useful tools to
obtain approximate values for these object that are not easily
derived by numerical methods. We end the paper with our
conclusions and comments in section \ref{Conclu} and an Appendix
where we discuss a useful representation for several integrals
appearing in the paper.

\section{Einstein-Rosen waves coupled to massless scalar matter}{\label{Eplusmatter}}

In this section we review in some detail the classical theory of
whole cylindrically symmetric spacetimes minimally coupled to a
cylindrically symmetric  massless scalar field. Although this
problem has been extensively considered in the literature (see for
instance \cite{Carmeli:1981nk} and references therein), it is
useful to gain some insight about the Hamiltonian formulation of
the model (that we will discuss in the next section) through the
analysis of the field equations and the Geroch formalism. As we
will see the solutions to the field equations strongly suggest
that the Hamiltonian of the system, obtained by carefully taking
into account surface terms in the action principle, is a simple
function of the Hamiltonians for two massless cylindrically
symmetric scalars. We start by considering the Einstein equations
for the system
\begin{eqnarray}
& & \square^{{\scriptscriptstyle(4)}}\phi=0\,,\label{EinsEqu1}
\\
&& R^{{\scriptscriptstyle(4)}}_{ab}=8\pi G_N\,
(\mathrm{d}\phi)_a(\mathrm{d}\phi)_b\,.\label{EinsEqu2}
\end{eqnarray}
Here $R^{{\scriptscriptstyle(4)}}_{ab}$ and
$\square^{{\scriptscriptstyle(4)}}$ are, respectively, the Ricci
tensor and the d'Alembert operator associated with the Levi-Civita
connection $D^{{\scriptscriptstyle(4)}}_a$ compatible with the
spacetime metric $g_{ab}^{{\scriptscriptstyle(4)}}$. The exterior
derivative of the matter scalar field $\phi$ is denoted by
$(\mathrm{d} \phi)_a$ and $G_N$ is the Newton constant.

\bigskip

Whole cylindrical symmetry \cite{Kuchar:1971xm} is characterized
by the existence of a $\mathbb{R}\times U(1)$ group of isometries
with two mutually orthogonal, hypersurface-orthogonal, commuting,
space-like globally defined Killing vectors $\xi^a$ and
$\sigma^a$. It is assumed also that the topology of the spacetime
is $\mathbb{R}^4$ and that the set of fixed points of the
rotations generated by $\sigma^a=(\partial/\partial \sigma)^a$
defines a 2-dimensional time-like surface: the axis of symmetry
$\{x\in\mathbb{R}^4\,|\,g_{ab}^{{\scriptscriptstyle(4)}}\sigma^a\sigma^b=0\}$.
On the other hand the translations along the symmetry axis
generated by $\xi^a=(\partial/\partial z)^a$ act freely and
satisfy the elementary flatness condition \cite{Carot:1999zm}.
This condition guarantees the $2\pi$ periodicity of the axial
coordinate $\sigma$. With these assumptions the Killing
coordinates $z\in \mathbb{R}$ and $\sigma\in[0,2\pi)$ are unique
up to the trivial transformations $z\mapsto \alpha z+ z_0$ and
$\sigma\mapsto \pm\sigma +\sigma_0$. In terms of these Killing
fields the symmetry of the system reflects on the vanishing of the
Lie derivatives
$$
L_\xi g^{{\scriptscriptstyle(4)}}_{ab}=L_\sigma
g^{{\scriptscriptstyle(4)}}_{ab}=0\,,\quad
L_\xi\phi=L_\sigma\phi=0\,.
$$
Finally, we will restrict our discussion to the class of
spacetimes for which the derivative $(\mathrm{d} R)_a$ of the
scalar field defined by
$$R^2=(g^{{\scriptscriptstyle(4)}}_{a_1a_2}\xi^{a_1}\xi^{a_2})
\,( g^{{\scriptscriptstyle(4)}}_{b_1b_2}\sigma^{b_1}\sigma^{b_2})
$$
is everywhere  space-like. $R$ is the area density of the isometry
group orbits.

\bigskip

In order to solve the Einstein equations (\ref{EinsEqu1},
\ref{EinsEqu2}) we will make use of the Geroch reduction technique
\cite{Geroch:1970nt}. First, owing to the fact that translations
have no fixed points, it is possible to rewrite the Einstein
equations as equations for fields defined on the quotient manifold
--topologically $\mathbb{R}^3$-- comprised by the translational
orbits. To do this we need to introduce a scalar field $\lambda$
and the tree dimensional spacetime metric
$g_{ab}^{{\scriptscriptstyle(3)}}$ given by\footnote{Here and in
the following $\xi_a=g^{{\scriptscriptstyle(4)}}_{ab}\xi^b$ and
$\sigma_a=g^{{\scriptscriptstyle(4)}}_{ab}\sigma^b$.}
\begin{eqnarray*}
\lambda= \xi_a\xi^a>0\,,\quad
g^{{\scriptscriptstyle(4)}}_{ab}=g^{{\scriptscriptstyle(3)}}_{ab}+\lambda^{-1}\xi_a\xi_b\,.
\end{eqnarray*}
These fields, as well as $\phi$, are well defined in the
translational orbit manifold. In terms of them the equations
(\ref{EinsEqu1}, \ref{EinsEqu2}) are equivalent to
\begin{eqnarray}
& & \square^{{\scriptscriptstyle(3)}}\phi
=-\frac{1}{2}g^{{\scriptscriptstyle(3)}ab}(\mathrm{d}\phi)_a
(\mathrm{d}\log\lambda)_b\,,\nonumber\\
& & \square^{{\scriptscriptstyle(3)}}
\lambda=\frac{1}{2\lambda}g^{{\scriptscriptstyle(3)}ab}
(\mathrm{d}\lambda)_a(\mathrm{d}\lambda)_b\,,
\label{EinsEqu3}\\
& & R_{ab}^{{\scriptscriptstyle(3)}}=\frac{1}{2\lambda}
D^{{\scriptscriptstyle(3)}}_a(\mathrm{d}\lambda)_b
-\frac{1}{4\lambda^2}(\mathrm{d}\lambda)_a(\mathrm{d}\lambda)_b+8\pi
G_N\, (\mathrm{d}\phi)_a(\mathrm{d}\phi)_b\,,\nonumber
\end{eqnarray}
where $R_{ab}^{{\scriptscriptstyle(3)}}$,
$D^{{\scriptscriptstyle(3)}}_a$, and
$\square^{{\scriptscriptstyle(3)}}$ refer to
$g^{{\scriptscriptstyle(3)}}_{ab}$. It is easy to prove that the
rotations generated by $\sigma^a$ are still a symmetry of the
reduced theory. In particular $\sigma^a$ is well defined in the
quotient manifold and satisfies
$$
L_\sigma g_{ab}^{{\scriptscriptstyle(3)}}=0\,, \quad L_\sigma
\lambda=0, \, \textrm{ and } \, L_\sigma \phi=0\,.
$$
At this point, although the rotations do not act freely, we can
still apply the Geroch reduction to the manifold obtained by
deleting the symmetry axis
$\rho=g_{ab}^{{\scriptscriptstyle(3)}}\sigma^a\sigma^b=0$ from
$\mathbb{R}^3$, and incorporating these points at the end of the
process by imposing regularity at the axis. In particular, outside
of the axis, we can define
\begin{eqnarray*}
\rho=g^{{\scriptscriptstyle(4)}}_{ab}\sigma^a\sigma^b=
g^{{\scriptscriptstyle(3)}}_{ab}\sigma^a\sigma^b>0\,\quad\mathrm{and}\quad
g^{{\scriptscriptstyle(3)}}_{ab}=g^{{\scriptscriptstyle(2)}}_{ab}
+\rho^{-1}\sigma_a\sigma_b\,.
\end{eqnarray*}
The equations (\ref{EinsEqu3}) can then be written as
\begin{eqnarray}
& &
\square^{{\scriptscriptstyle(2)}}\phi=-\frac{1}{2}g^{{\scriptscriptstyle(2)}ab}
(\mathrm{d}\phi)_a(\mathrm{d}\log(\lambda\rho))_b \nonumber\\
& & \square^{{\scriptscriptstyle(2)}}\lambda=
-\frac{\lambda}{2}g^{{\scriptscriptstyle(2)}ab}
(\mathrm{d}\log\lambda)_a(\mathrm{d}\log(\lambda^{-1}\rho))_b
\nonumber
\\
& & \square^{{\scriptscriptstyle(2)}}\rho=
\frac{\rho}{2}g^{{\scriptscriptstyle(2)}ab}(\mathrm{d}\log\rho)_a
(\mathrm{d}\log(\lambda^{-1}\rho))_b\nonumber \\
& &R_{ab}^{{\scriptscriptstyle(2)}}=\frac{1}{2\lambda}
D^{{\scriptscriptstyle(2)}}_a(\mathrm{d}\lambda)_b+\frac{1}{2\rho}
D^{{\scriptscriptstyle(2)}}_a(\mathrm{d}\rho)_b
-\frac{1}{4}(\mathrm{d}\log\lambda)_a(\mathrm{d}\log\lambda)_b
-\frac{1}{4}(\mathrm{d}\log\rho)_a(\mathrm{d}\log\rho)_b+8\pi G_N
(\mathrm{d}\phi)_a (\mathrm{d}\phi)_b \,,\nonumber
\end{eqnarray}
where the notation in the previous expressions is the natural one.
These 2-dimensional field equations can be solved in two steps.
First, it is convenient to replace the field $\rho$ in terms of
$R=\sqrt{\lambda\rho}$ and $\lambda$ to get
\begin{eqnarray}
& & \square^{{\scriptscriptstyle(2)}}\phi+
g^{{\scriptscriptstyle(2)}ab}(\mathrm{d}\phi)_a(\mathrm{d}\log R)_b=0 \nonumber \\
& & \square^{{\scriptscriptstyle(2)}}\log\lambda+
g^{{\scriptscriptstyle(2)}ab}(\mathrm{d}\log\lambda)(\mathrm{d}\log R)=0 \label{EinsEqu4}\\
& & \square^{{\scriptscriptstyle(2)}} R=0 \nonumber\\
& &
R_{ab}^{{\scriptscriptstyle(2)}}=\frac{1}{R}D^{{\scriptscriptstyle(2)}}_a(\mathrm{d}R)_b
-\frac{1}{2} [\mathrm{d}\log\lambda]_{(a}
[\mathrm{d}\log(\lambda^{-1}R^2)]_{b)}+8\pi G_N\,
(\mathrm{d}\phi)_a (\mathrm{d}\phi)_b\,.\nonumber
\end{eqnarray}
Second, owing to the fact that we are dealing now with
2-dimensional field equations,
$\square^{{\scriptscriptstyle(2)}}R=0$ allows us to introduce a
new scalar field $T$, the harmonic function conjugate to $R$,
whose derivative $(\mathrm{d} T)_a$ is everywhere time-like by
means of
$$(\mathrm{d}T)_a=
\epsilon^{{\scriptscriptstyle(2)}}_{ab}g^{{\scriptscriptstyle(2)}bc}
(\mathrm{d} R)_c\,,$$ where
$\epsilon^{{\scriptscriptstyle(2)}}_{ab}$ is the volume element
associated to $g^{{\scriptscriptstyle(2)}}_{ab}$. Notice that this
definition is conformally invariant. It is now possible to use $R$
and $T$ as coordinates in the 1+1 reduced spacetime and introduce
the flat metric
\begin{eqnarray*}
\eta_{ab}^{{\scriptscriptstyle(2)}}=-(\mathrm{d}T)_a(\mathrm{d}T)_b
+(\mathrm{d}R)_a(\mathrm{d}R)_b\,.
\end{eqnarray*}
The  degrees of freedom of $g_{ab}^{{\scriptscriptstyle(2)}}$ are
encoded in the conformal factor $e^{\gamma}$ defined by
\begin{eqnarray*}
g_{ab}^{{\scriptscriptstyle(2)}}=\frac{e^\gamma}{\lambda}
\,\eta_{ab}^{{\scriptscriptstyle(2)}}\,.
\end{eqnarray*}
In terms of these new fields, the equations (\ref{EinsEqu4})
become
\begin{eqnarray*}
& & \square\phi+
\eta^{{\scriptscriptstyle(2)} ab}(\mathrm{d} \phi)_a(\mathrm{d} \log R)_b=0\\
& & \square\log\lambda+\eta^{{\scriptscriptstyle(2)}
ab}(\mathrm{d} \log
\lambda)_a (\mathrm{d} \log R)_b=0\\
& &
\eta^{{\scriptscriptstyle(2)}cd}
\big[\frac{1}{2}(\mathrm{d}\log\lambda)_c(\mathrm{d}
\log\lambda)_d+8\pi G_N\, (\mathrm{d}\phi)_c(\mathrm{d} \phi)_d-
(\mathrm{d}\gamma)_c(\mathrm{d}\log
R)_d\big]\,\eta_{ab}^{{\scriptscriptstyle(2)}}\\
& &\hspace{6cm}=\frac{2}{R}\partial_a(\mathrm{d}
R)_b-2(\mathrm{d}\gamma)_{(a}(\mathrm{d}\log R)_{b)} +(\mathrm{d}
\log\lambda)_a(\mathrm{d}\log\lambda)_b+8\pi
G_N\,(\mathrm{d}\phi)_a(\mathrm{d}\phi)_b,
\end{eqnarray*}
where $\square$ is the d'Alembert operator defined by the flat
metric $\eta^{{\scriptscriptstyle(2)}}_{ab}$. The general solution
to these equations can be written in a very convenient form in
terms of the fields
\begin{eqnarray*}
\phi_g:=\log\lambda\,,\quad \phi_s:=\sqrt{16\pi G_N}\phi\,.
\end{eqnarray*}
By using $R$ and $T$ as coordinates, the Einstein equations are
equivalent to two uncoupled cylindrically symmetric Klein-Gordon
equations
\begin{eqnarray*}
&&[-\partial^2_T +\partial^2_R+\frac{1}{R}\partial_R] \phi_g=0\,,\\
&&[-\partial^2_T +\partial^2_R+\frac{1}{R}\partial_R] \phi_s=0\,,\\
&&\partial_T\gamma=R[(\partial_R\phi_g)(\partial_T\phi_g)
+(\partial_R\phi_s)(\partial_T\phi_s)]\,,\\
&&\partial_R\gamma=\frac{R}{2}[(\partial_T\phi_g)^2+(\partial_R\phi_g)^2+
(\partial_T\phi_s)^2+(\partial_R\phi_s)^2].
\end{eqnarray*}
The first two are equations for a massless, axially symmetric,
scalar fields in a Minkowskian background and the equations for
$\gamma$ satisfy an integrability condition that allows us to
immediately write their solution as
\begin{eqnarray*}
\gamma=\frac{1}{2}\int [(\partial_T\phi_g)^2+(\partial_R\phi_g)^2+
(\partial_T\phi_s)^2+(\partial_R\phi_s)^2]\,R\,\mathrm{d}R.
\end{eqnarray*}
Finally the four-dimensional spacetime metric satisfying the
Einstein field equations can be written as
\begin{eqnarray*}
g_{ab}^{{\scriptscriptstyle(4)}}=e^{\gamma-\phi_g}\big[-(\mathrm{d}T)_a(\mathrm{d}T)_b
+(\mathrm{d}R)_a(\mathrm{d}R)_b\big]
+R^2e^{-\phi_g}(\mathrm{d}\sigma)_a(\mathrm{d}\sigma)_b
+e^{\phi_g}(\mathrm{d}z)_a(\mathrm{d}z)_b\,.
\end{eqnarray*}
The form for the `C-energy' $\gamma$ strongly suggests that the
Hamiltonian of the system can be obtained from the one
corresponding to Einstein-Rosen waves by adding the contribution
of the extra scalar field. We show this in the next section.

\section{Hamiltonian Formalism}{\label{Hamil}}

In order to develop the Hamiltonian formalism, we start from the
Einstein-Hilbert action in four-dimensions for gravity coupled to
a massless cylindrically symmetric scalar $\phi_s$:
\begin{eqnarray*}
S^{{\scriptscriptstyle(4)}}=\frac{1}{16\pi
G_N}\int_{\mathcal{M}^3\times
Z}\,|g^{{\scriptscriptstyle(4)}}|^{1/2}\bigg(R^{{\scriptscriptstyle(4)}}
-\frac{1}{2}g^{{\scriptscriptstyle(4)}ab}(\mathrm{d}\phi_s)_a
(\mathrm{d}\phi_s)_b\bigg) +\frac{1}{8\pi
G_N}\int_{\partial(\mathcal{M}^3\times
Z)}\,\big(|h^{{\scriptscriptstyle(3)}}|^{1/2}K-
|h^{{\scriptscriptstyle(3)0}}|^{1/2}K^0\big)\,.
\end{eqnarray*}
We have included the boundary terms needed to have a well defined
variational principle. The fields are taken to be regular in the
symmetry axis and the boundary conditions at infinity correspond
to the definition of asymptotic flatness introduced by Ashtekar
and Varadarajan in the 2+1 dimensional setting
\cite{Ashtekar:1994ds}. We use here the rescaled scalar $\phi_s$
introduced above with the same dimensions of $\phi_g$. As we will
see they play symmetric roles in the final formulation of the
model. The four-dimensional manifold where the previous action is
defined has the form of a product $\mathcal{M}^3\times Z$ where
$\mathcal{M}^3$ is a three-dimensional manifold orthogonal to the
translational Killing vector $\xi^{a}=(\partial/\partial z)^a$ and
$Z=[z_1,z_2]$ is a closed interval in this direction (axis of
symmetry). We have introduced also a fiducial metric
$g^{{\scriptscriptstyle(4)0}}_{ab}$ that provides us with an
origin for the energy, ensures that the action is finite and fixes
the asymptotic behavior of the fields in such a way that the
Minkowski metric has zero energy. Finally
$h^{{\scriptscriptstyle(3)}}_{ab}$ and
$h^{{\scriptscriptstyle(3)0}}_{ab}$ are the induced metrics on the
boundary. Owing to the translation symmetry it is possible to
rewrite the previous action as an equivalent one in 2+1 dimensions
that can be interpreted as the Einstein-Hilbert action with two
massless scalars after the conformal transformation
$g_{ab}=e^{\phi_g} g^{{\scriptscriptstyle(3)}}_{ab}$ is performed.
Hence, our starting point to get the Hamiltonian will be
\begin{eqnarray}
S^{{\scriptscriptstyle(3)}}=\frac{1}{16\pi
G_3}\int_{\mathcal{M}^3}\,|g|^{1/2}\bigg(\,
R^{{\scriptscriptstyle(3)}}-\frac{1}{2}g^{ab}(\mathrm{d}\phi_g)_a(\mathrm{d}\phi_g)_b
-\frac{1}{2}g^{ab}(\mathrm{d}\phi_s)_a(\mathrm{d}\phi_s)_b\bigg)+\frac{1}{8\pi
G_3}\int_{\partial\mathcal{M}^3}\,\big(|h|^{1/2}K-|h^0|^{1/2}K^0\big)\,.\label{001}
\end{eqnarray}
Here all the geometrical objects refer to the metric $g_{ab}$. The
coupling constant $G_3$ is the gravitational constant per unit
length along the symmetry axis and in the following we choose
units such that $c=1$. In this three dimensional expression of the
action, we notice that the scalar field term plays exactly the
same role as the gravitational scalar. It is important to point
out that both fields are coupled through the metric, but not
directly (there are no cross terms in the action).

To obtain the Hamiltonian we follow the procedure developed in
\cite{Ashtekar:1996bb} for the vacuum case. First of all, we
choose a foliation of $\mathcal{M}^3$ with timelike unit normal
$n^{a}$, a radial unit vector $\hat{r}^a$, and denote as
$\sigma^a$ the azimutal, hypersurface orthogonal, Killing vector
field (notice that this is not a unit vector). It is possible now
to write the metric as
$g_{ab}=-n_an_b+\hat{r}_a\hat{r}_b+\frac{1}{R^2}\sigma_a\sigma_b$
(with $R^2= g_{ab}\sigma^a\sigma^b$). We also introduce two
additional vector fields $t^{a}$ and $r^{a}$ defined as $t^a=N
n^a\!+\!N^r \hat{r}^a$ and $r^a=e^{\gamma/2}\hat{r}^a$, where $N$
is the lapse function, $N^r$ the radial shift, and at this point
$\gamma$ is just an extra field (that will eventually coincide
with the one introduced in the previous section). We impose the
condition that the commutators of these new vector fields are
zero, so we can define coordinates $t,r,\theta$ and get the
following consistency conditions
\begin{eqnarray*}
&&\partial_{\sigma}N=\partial_{\sigma}N^r=\partial_{\sigma}\gamma=0\,;
\,[\sigma,\hat{r}]^a=[\sigma,n]^a=0\,;\\
&& n^a\partial_r N+\hat{r}^a(\partial_rN^r-\partial_t
e^{\gamma/2})+Ne^{\gamma/2}[\hat{r},n]^a=0.
\end{eqnarray*}
Finally the metric takes the form:
\begin{eqnarray*}
g_{ab}=(N^{r2}-N^2)(\mathrm{d}t)_a(\mathrm{d}t)_b
+2e^{\gamma/2}N^r(\mathrm{d}t)_{(a}(\mathrm{d}r)_{b)}
+e^{\gamma}(\mathrm{d}r)_a(\mathrm{d}r)_b
+R^2(\mathrm{d}\sigma)_a(\mathrm{d}\sigma)_b.
\end{eqnarray*}
Now, if we take the boundary $\partial\mathcal{M}^3$ to be
orthogonal to the vectors $\hat{r}^a$ and $n^{a}$, it is
straightforward to give an expression for the action in three
dimensions in terms of the fields $N$, $N^r$, $\gamma$, $R$, and
$\phi_{g,s}$:
\begin{eqnarray*}
S^{{\scriptscriptstyle(3)}}&=&\frac{1}{8
G_3}\int_{t_1}^{t_2}\!\!\int_0^{\tilde{r}}\!\!
\bigg(Ne^{-\gamma/2}(\gamma^{\prime}R^{\prime}-2R^{\prime\prime})
-\frac{1}{N}(e^{\gamma/2}\dot{\gamma}-2N^{r\prime})
(\dot{R}-e^{-\gamma/2}N^rR^{\prime})\\
&+&\frac{R}{2N}\Big[e^{\gamma/2}\dot{\phi}^2_g-2N^r\dot{\phi}_g{\phi}_g^{\prime}+
e^{-\gamma/2}(N^{r2}-N^2){\phi}_g^{\prime2}\Big]
+\frac{R}{2N}\Big[e^{\gamma/2}\dot{\phi}^2_s-2N^r\dot{\phi}_s{\phi}_s^{\prime}+
e^{-\gamma/2}(N^{r2}-N^2){\phi}_s^{\prime2}\Big]\bigg)\,\mathrm{d}r\,\mathrm{d}t
\\&+&\frac{1}{4
G_3}\int_{t_1}^{t_2}\!\!(Ne^{-\gamma/2}R^{\prime}-1)\,\mathrm{d}t\,,
\label{006}
\end{eqnarray*}
where we have denoted $\partial_t$ with a dot and $\partial_r$
with a prime. We get now the Hamiltonian for the special case in
which the boundary is taken to infinity
($\tilde{r}\rightarrow\infty$) taking into account that the metric
$g_{ab}$ reduces to Minkowskian metric when $N=1$, $N^r=0$,
$\gamma=0$, and $R=r$, assuming regularity in the axis, and the
2+1 dimensional asymptotic flatness conditions for the fields
introduced in \cite{Ashtekar:1994ds,Ashtekar:1996bb}. The
Hamiltonian is then
\begin{eqnarray*}
H&=&\int_0^{\infty}
\bigg(N^re^{-\gamma/2}\Big[p_RR^{\prime}-2p^{\prime}_{\gamma}+p_{\gamma}\gamma^{\prime}+
\phi^{\prime}_gp_g+\phi^{\prime}_sp_s\Big] \\
&&\hspace{7mm} +Ne^{-\gamma/2} \Big[\frac{1}{8
G_3}(2R^{\prime\prime}-\gamma^{\prime}R^{\prime})
-8G_3p_Rp_{\gamma}+\frac{4G_3}{R}p_g^2+\frac{R}{16G_3}\phi_g^{\prime2}+
\frac{4G_3}{R}p_s^2+\frac{R}{16G_3}\phi_s^{\prime2}\Big]\bigg)\mathrm{d}r
\\&+&\frac{1}{4G_3}(1-e^{-\gamma_{\infty}/2})\,,\label{008}
\end{eqnarray*}
where $p_R$, $p_{\gamma}$, $p_g$, and $p_s$ are the momenta
canonically conjugate to $R$, $\gamma$, $\phi_g$, and $\phi_s$
respectively and $\gamma_{\infty}:=
\lim_{\tilde{r}\rightarrow\infty}\gamma(\tilde{r})$. It is easy to
read both the constraints and the reduced Hamiltonian from the
last expression. In order to proceed further we fix the gauge with
the same conditions as in the absence of matter
\cite{Ashtekar:1996bb}
$$R(r)=r \quad\textrm{ and }\quad p_{\gamma}(r)=0\,;$$
it is straightforward to show that these gauge fixing conditions
are admissible. We can now solve the constraints to get
\begin{eqnarray*}
\gamma(\tilde{R})&=&\frac{1}{2}\int_0^{\tilde{R}}
\Big(\phi_g^{\prime2}+\frac{(8G_3p_g)^2}{R^2}+\phi_s^{\prime2}
+\frac{(8G_3p_s)^2}{R^2}\Big)\, R \mathrm{d}R\,,
\\
p_R&=&-p_s \phi'_s-p_g \phi'_g\,.
\end{eqnarray*}
Finally, the three-dimensional line element can be written as
\begin{equation}
ds^2=e^{\gamma}[-e^{-\gamma_{\infty}}dt^2+dR^2]+R^2d\sigma^2\,,\label{009}
\end{equation}
the reduced phase space is coordinatized by $\phi_s(R)$, $p_s(R)$,
$\phi_g(R)$, and $p_g(R)$; and the reduced Hamiltonian is
$$H=\frac{1}{4G_3}(1-e^{-\gamma_{\infty}/2})$$
where
$$\gamma_{\infty}=\frac{1}{2}\int_0^{\infty}
\Big(\phi_g^{\prime2}+\frac{(8G_3p_g)^2}{R^2}+\phi_s^{\prime2}
+\frac{(8G_3p_s)^2}{R^2}\Big)\, R \mathrm{d}R.$$ As we can see
$\gamma_{\infty}$ is the Hamiltonian for a system of two free,
axially symmetric scalar fields in 2+1 dimensions and the true
Hamiltonian $H$ is a non-linear and bounded \textit{function} of
this free Hamiltonian, similar to the one appearing in the absence
of matter \cite{Ashtekar:1994ds}. The Hamilton equations are
\begin{eqnarray*}
\dot{\phi}_{s,g}=e^{-\gamma_\infty/2}\,\frac{p_{s,g}}{R}\,,\quad
\dot{p}_{s,g}=e^{-\gamma_\infty/2}(R\phi_{s,g}^{\prime})'\,.
\end{eqnarray*}
Though they are non-linear integro-differential equations they can
be easily solved by realizing that $\gamma_{\infty}$ is a constant
of motion. Taking this fact into account, we can perform a change
in the time coordinate $T=e^{-\gamma_{\infty}/2}\,t$ and rewrite
them as
\begin{eqnarray}
&&\hspace{-1cm}[-\partial^2_{T}+\partial_R^2+\frac{1}{R}\partial_R]\,\phi_{s,g}=0;
\label{011}
\end{eqnarray}
describing two massless, axially symmetric scalar fields in 2+1
dimensional Minkowskian background. As we see this change in the
time variable provides a one to one map from the solutions to a
simple linear system to those of the non-linear one. This mapping
encodes the non-trivial interaction present in the model. If we
compare the classical evolution of two different sets of initial
data we can see that they both correspond to the evolution defined
by the free Hamiltonian $\gamma_{\infty}$ with times elapsing at
different rates (defined by the conserved values of
$\gamma_{\infty}$). Notice that once we have a particular solution
for equations (\ref{011}) we have the freedom to ``change
coordinates" in the metric and write it in terms of $t$ or in
terms of $T$. This is not the case quantum mechanically, because
the evolution of arbitrary states involves, in general, the
\textit{superposition} of Hilbert space vectors with energy (and
time) dependent phases. This implies that quantum dynamics will be
much more complicated than the classical one. In conclusion,  the
fact that the Hamiltonian is a function of a certain free
Hamiltonian makes it both non trivial (we are, indeed, dealing
with a coupled system) and solvable.

\section{Canonical Quantization}{\label{quant}}

Once we have characterized the reduced phase space and obtained
the classical Hamiltonian, we proceed to quantize the model. To
this end we will use the Fock Hilbert spaces $\mathcal{F}_{g,s}$
associated with two different free, massless, axially symmetric
scalar fields propagating in a Minkowskian background. These
spaces are endowed with the usual creation and annihilation
operators $\hat{a}_{g,s}(k),\hat{a}^{\dagger}_{g,s}(k)$ satisfying
\begin{eqnarray*}
&&[\hat{a}_g(k),\hat{a}^{\dagger}_g(q)]=\delta(k,q),
\quad[\hat{a}_s(k),\hat{a}^{\dagger}_s(q)]=\delta(k,q);\quad\quad
\end{eqnarray*} their corresponding vacua are denoted as $|0\rangle^{g,s}$.
The Hilbert space of the interacting model is taken as the tensor
product $\mathcal{H}=\mathcal{F}_g\otimes \mathcal{F}_s$ of the
Fock spaces corresponding to both scalars. We define the
annihilation operators\footnote{Creation operators are defined in
an analogous way.} for modes of ``gravitational" or ``matter"
types as $\hat{A}_{g}(k):=\hat{a}_{g}(k)\otimes\mathbb{I}_s$,
$\hat{A}_{s}(k):= \mathbb{I}_g\otimes \hat{a}_{s}(k)$, and the
distribution-valued, field and momentum operators
$\hat{\phi}_{g,s}(R)$, $\hat{p}_{g,s}(R)$
\begin{eqnarray*}
&& \hspace{-6mm}\hat{\phi}_{g,s}(R)=\sqrt{4
G_3\hbar}\!\int_0^{\infty}
J_0(Rk)\,[\hat{A}_{g,s}(k)+\hat{A}_{g,s}^{\dagger}(k)]\,\mathrm{d}k\,,\\
&&
\hspace{-6mm}\hat{p}_{g,s}(R)=\frac{iR}{2}\sqrt{\frac{\hbar}{4G_3}}\int_0^{\infty}
kJ_0(Rk)\,[\hat{A}_{g,s}^{\dagger}(k)-\hat{A}_{g,s}(k)]\,\mathrm{d}k\,,
\end{eqnarray*}
satisfying the usual commutation relations
$[\hat{\phi}_{g,s}(R_1),\hat{p}_{g,s}(R_2)]=i\hbar\delta(R_1,R_2)$.
Notice that we can construct states with a fixed number of quanta
of ``gravitational" or ``scalar" type by acting with the
corresponding creation operators  on the vacuum state
$|\Omega\rangle=|0\rangle^g\otimes|0\rangle^s\in\mathcal{H}$ that
is the minimum energy eigenstate of the quantum
Hamiltonian\footnote{We have normal ordered the exponent because,
otherwise, the Hamiltonian is trivial.}
\begin{widetext}
\begin{eqnarray}
\hat{H}\!=\!\frac{1}{4G_3}\Bigg[1\!-\!\exp\Big(-\!4G_3\hbar\!\displaystyle
\int_0^{\infty}\!\!k\,[\hat{A}_g^{\dagger}(k)\hat{A}_g(k)\!
+\!\hat{A}_s^{\dagger}(k)\hat{A}_s(k)]\,\mathrm{d}k\Big)\!\Bigg].\label{HamG}
\end{eqnarray}
\end{widetext}
This quantum Hamiltonian is a nonlinear and bounded function of
the sum of the free Hamiltonians
$$
\hat{H}_0^{g,s}=\int_0^{\infty}\!\!k\,
\hat{A}_{g,s}^{\dagger}(k)\hat{A}_{g,s}(k)\,\mathrm{d}k
$$
for two massless, cylindrically symmetric scalar fields in 2+1
dimensions evolving in a fictitious Minkowskian background. Their
sum $\hat{H}_0^g+\hat{H}_0^s$ is an observable but it is not the
generator of the time evolution of the system. The physical
evolution, from $t_0$ to $t$, is generated by $\hat{H}$ and is
given by the unitary evolution operator
\begin{eqnarray}
\hat{U}(t,t_0)=\exp\Big(-\frac{i(t-t_0)}{4G_3\hbar}\Big[
1-e^{-4G_3\hbar(\hat{H}_0^g+\hat{H}_0^s)}\Big]\Big).\label{016}
\end{eqnarray}
This operator defines the $S$ matrix of the system when we take
the appropriate time limits. Its matrix elements on $n$-particle
states are straightforward to compute because these are
eigenstates of the free Hamiltonian
$\hat{H}_0=\hat{H}_0^g+\hat{H}_0^s$. As we can see the only matrix
elements --involving state vectors with a definite number of both
type of quanta-- that are non zero are those connecting states
with the same number of particles of each type; hence there is no
conversion of quanta of one type into the other. At this point, it
is important to reflect upon the interpretation of these
elementary excitations of the fields. One must be careful, for
example, when interpreting states such as
$|0\rangle^g\otimes|\Phi\rangle^s$ because one should not be lead
to think of them as matter (represented by $|\Phi\rangle^s$)
evolving in a certain background (given by $|0\rangle^g$). The
system that we are considering is a coupled one, and hence one is
not entitled to think of the metric and matter fields as
independent objects. They are coupled in the equations and, in
particular, the classical metric depends on both the gravitational
and matter fields. Conversely, the evolution of the scalar field
depends on the metric. If we want to approximate the Minkowski
metric, we have that the state that most closely resembles it is
the vacuum of our total Hilbert space $|\Omega\rangle$ [by the
way, this is the only coherent state of the system that we know
under the evolution (\ref{016})]. In the next sections we will
make extensive use of states of the type
$|0\rangle^g\otimes|k\rangle^s$ consisting of tensor products of
the vacuum state of one of the Fock spaces and a one particle
state of the other. As we will show in the last sections of the
paper these single particle states are the closest ones to
Minkowski if one wants to incorporate an extra element that can be
used to explore and describe the geometry of the quantized model
in an operational way.

\bigskip

In the following it will be convenient to explicitly keep the
length scale of the system $G=G_3\hbar$ in the mathematical
expressions of the relevant objects. We will nevertheless use
units such as $\hbar=1$. With the time evolution (\ref{016})
 defined by the Hamiltonian (\ref{HamG}) the annihilation and
creation operators in the Heisenberg picture are
\begin{eqnarray}
&&\hat{A}_{s,g}(k;t,t_0)=\hat{U}^\dagger(t,t_0)\hat{A}_{s,g}(k)\hat{U}(t,t_0)=
\exp\!\left[-i(t-t_0)E(k)e^{-4G\hat{H}_0}\right]\!\hat{A}_{s,g}(k),\nonumber\\
&&\hat{A}_{s,g}^\dagger(k;t,t_0)=\hat{U}^\dagger(t,t_0)
\hat{A}_{s,g}(k)\hat{U}(t,t_0)=\hat{A}_{s,g}^\dagger(k)\,\exp\!\left[i(t-t_0)E(k)
e^{-4G\hat{H}_0}\right],\nonumber
\end{eqnarray}
where $$E(k):=\frac{1}{4G}(1-e^{-4Gk})\quad \textrm{ and }\quad
\hat{H}_0:=\hat{H}_0^g+\hat{H}_0^s\,.$$ Then the scalar field
operators that describe the gravitational and the massless scalar
field degrees of freedom at time $t$  are
\begin{widetext}
\begin{eqnarray*}
\hat{\phi}_{s,g}(R;t,t_0)=\sqrt{4G}\int_0^\infty\!\!\!
J_0(Rk)\left[\,\hat{A}_{s,g}(k;t,t_0)+\hat{A}_{s,g}^\dagger(k;t,t_0)\,\right]
\mathrm{d}k\,.
\end{eqnarray*}

\section{Two-point functions}{\label{two}}

As commented in the introduction one of the main goals of the
paper is to find a way to recover a physical picture of spacetime
in the quantized symmetry reduction of gravity coupled to matter
discussed above. We want to find ways to describe a quantized
spacetime geometry in an operational way much in the same way as
one explores a classical spacetime geometry by using test
particles. To this end it is useful to have objects playing the
role of particle propagators or, even better, one-particle states
that could allow us to define suitable wave functions with a
straightforward interpretation as spatial probability amplitudes.
Hopefully the time evolution of these objects may give us some
idea about the physical effects of quantizing the gravitational
field and also tell us something about how the classical
macroscopic geometry emerges. As the reader may expect this is not
easy; in fact the problem of finding suitable \textit{position}
eigenstates in the usual Minkowskian QFT's is already non-trivial.
Our strategy will be to use some of the objects introduced in the
discussion of these issues in the traditional approaches to QFT
(such as two-point functions or Newton-Wigner states), interpret
them in our framework, and use them to obtain a physical picture
of the quantized geometry and gravity. Also, in order to
disentangle genuine quantum gravitational phenomena from artifacts
introduced by the symmetry of the problem and the reduction
process we will compare the relevant objects to the corresponding
ones in a model of a quantized, axially symmetric, massless free
scalar field moving in a Minkowskian background. This comparison
is conceptually simpler if one works both with gravity and the
scalar field.

\bigskip

The vacuum expectation values of the product of two fields at
different spacetime points can be interpreted, at least in an
approximate sense, as propagation amplitudes for particles or
field quanta created at a certain event to be found at another.
This is so because the (Schr\"odinger picture) scalar field
operators that describe the gravitational and scalar degrees of
freedom are
\begin{widetext}
\begin{eqnarray*}
\hat{\phi}_{s,g}(R)=\sqrt{4G}\int_0^\infty\!\!\!
\,J_0(Rk)\left[\hat{A}_{s,g}(k)+\hat{A}_{s,g}^\dagger(k)\right]\mathrm{d}k\,
\end{eqnarray*}
and their action on the vacuum
$|\Omega\rangle=|0\rangle^g\otimes|0\rangle^s$ satisfies
$$
\frac{1}{\sqrt{4G}}\,\hat{\phi}_{s,g}(R)|\Omega\rangle=\int_0^\infty\!\!\!\mathrm{d}k\,
\,J_0(Rk)\hat{A}_{s,g}^\dagger(k)|\Omega\rangle=\int_0^\infty\!\!\!\mathrm{d}k\,
\,J_0(Rk)|k\rangle_{s,g}.
$$
\end{widetext}
These are linear superpositions of (orthonormal) states
$|k\rangle_{s,g}:= \hat{A}_{g,s}^{\dagger}(k)|\Omega\rangle$ with
well-defined ``radial momentum" $k$. Notice that
\begin{equation}
J_0(Rk)=\frac{1}{\sqrt{4G}} {_{\,s,g\!}\langle}
k|\hat{\phi}_{s,g}(R)|\Omega\rangle\label{radialpsi}
\end{equation}
is a solution of the radial part of the Schr\"odinger equations
for states with zero angular momentum in two dimensions
$$
[\partial_R^2+\frac{1}{R}\partial_R+k^2]J_0(Rk)=0.
$$
If we consider a small volume element $\Delta V$ at a distance $R$
from the symmetry axis the value of $J^2_0(Rk)\Delta V$ is
proportional to the probability of finding a particle of type $s$
or $g$ inside it. Notice that \textit{this is not} in general the
probability to find the particle in a thin cylindrical shell of
radius $R$.

In order to consider the amplitude for propagation of quanta of
the matter scalar we could consider more general situations, for
example we might take states of the form
$|c\rangle^g\!\otimes|k\rangle^s$ with $|c\rangle^g$ a suitable
``coherent" gravitational state under the evolution defined by the
dynamics of the system. Notice, however, that the physical
interpretation of such a state is not completely clear and, in
particular, one should not be led to think that the gravitational
part of the state (say $|c\rangle^g$) fixes the geometry (or a
suitable classical approximation thereof) and the matter part the
scalar field; in fact \textit{both} parts of the state vector
contribute to fix the metric on one hand and the matter field on
the other (classically this can be understood by realizing that
the metric depends on \textit{both} the gravitational and matter
scalars).

\bigskip

In the following we will consider the case $t_2>t_1$ and
interpret the  matrix element
$\langle\Omega|\hat{\phi}_{s,g}(R_2;t_2,t_0)\hat{\phi}_{s,g}(R_1;t_1,t_0)
|\Omega\rangle$ as the (approximate) probability amplitude of a
particle created at a point at a distance $R_1$ from the axis in
the instant of time $t_1$ to be detected at a another point at
$R_2$ distance in the instant of time $t_2$. We can now obtain in
a straightforward way\footnote{The creation and annihilation
operators as written above are specially suitable for this
computation and similar ones where the states are eigenstates of
the Hamiltonian operator.}
\begin{eqnarray}
\langle\Omega|\hat{\phi}_{s,g}(R_2;t_2,t_0)\hat{\phi}_{s,g}(R_1;t_1,t_0)
|\Omega\rangle\!\!&=&\!\!4G
\int_0^{\infty}\!\!\!\,J_0(R_1k)J_0(R_2k)\langle\Omega|\hat{A}_{s,g}(k;t_2,t_0)
\hat{A}_{s,g}^\dagger(k;t_1,t_0)|\Omega\rangle\, \mathrm{d}k
\nonumber\\
&=&\!\!4G\int_0^{\infty} \!\!\!
J_0(R_1k)J_0(R_2k)\exp[-i(t_2-t_1)E(k)]\,\mathrm{d}k.\label{f1}
\end{eqnarray}
\end{widetext}
We have introduced an initial time $t_0$ that will not appear in
the final expressions of the matrix elements that we will consider
here so that in the following we will write
$\langle\Omega|\hat{\phi}_{s,g}(R_2,t_2)\hat{\phi}_{s,g}(R_1,t_1)|\Omega\rangle$.

Let us consider the integral (\ref{f1}). First of all it must be
said that it is not possible to compute it in closed form although
there are suitable ways to compute it numerically and approximate
it by means of asymptotic expansions \cite{BarberoG:2004uc}. The
relevant parameters in the integral are: $R_1$, $R_2$, $t_2-t_1$
--the arguments of the two-point function--, and $4G$ that sets
the length scale.\footnote{In fact $4G$ plays the role of the
Planck length. Notice also that it also sets the time and energy
scales as we are taking units such that $\hbar=c=1$.} In view of
this it appears to be appropriate to refer both length and time to
this scale and introduce the adimensional variables
$\rho_1=\frac{R_1}{4G}$, $\rho_2=\frac{R_2}{4G}$, and
$\tau=\frac{t_2-t_1}{4G}$ together with the change of variables
$q=4Gk$ that gives a dimensionless integration variable. In this
way we can rewrite (\ref{f1}) as
\begin{equation}
\langle\Omega|\hat{\phi}_{s,g}(R_2,t_2)
\hat{\phi}_{s,g}(R_1,t_1)|\Omega\rangle=\int_0^{\infty}\!\!\!
J_0(\rho_1q)J_0(\rho_2q)\exp[-i\tau
(1-e^{-q})]\,\mathrm{d}q\,,\label{f4}
\end{equation}
where $\rho_1$ and $\rho_2$ are to be considered as functions of
$R_1$ and $R_2$ as defined above. An interesting consequence of
(\ref{f4}) is that we can obtain the vacuum expectation value of
the commutator of Heisenberg picture field operators by taking its
imaginary part (as we did in \cite{BarberoG:2004uc} to discuss the
microcausality in this system).

One can consider, in principle, the numerical computation of this
type of improper integral but this is rather difficult due to the
oscillating nature of the integrand. In spite of this there are
efficient ways to do it, as the one described in the appendix, by
rewriting it as an integral over a torus plus a rapidly convergent
improper integral. On the other hand, the advantage of using
asymptotic approximations in some relevant parameters lies on the
fact that it gives the limiting behavior at large scales in some
physically relevant regimes and also allows to get numerical
estimates in a rapid manner. It is possible to consider separate
expansions for each of the parameters $\rho_1$, $\rho_2$, and
$\tau$ which are valuable in the sense that it is possible to
study the behavior of the two-point function when only one of them
is taken to be large. It is also possible to write down an
expansion in which all of them are simultaneously large while
keeping their relative values. Here this approximation corresponds
to large length and time intervals as compared to the scale set by
$G$. Let us consider these separately:

\subsection{Asymptotic expansions in $\rho_1$ or $\rho_2$}

For large values of $\rho_1$ and $\rho_2$ the asymptotic behaviors
of (\ref{f4}) are respectively
$$
\frac{1}{\rho_1}+\frac{1}{\rho_1^3}\left[\frac{\rho_2^2}{4}+\tau^2
-\frac{i\tau}{2}\right]+O(\rho_1^{-6})
\quad \mathrm{and}\quad
\frac{1}{\rho_2}+\frac{1}{\rho_2^3}\left[\frac{\rho_1^2}{4}+\tau^2
-\frac{i\tau}{2}\right]+O(\rho_2^{-6}),
$$
they can be obtained in a straightforward way by considering
(\ref{f4}) as a standard $h$-transform with asymptotic parameters
$\rho_1$ or $\rho_2$ and using Mellin transform techniques
\cite{Handels}. The imaginary part of the previous expressions
corresponds to the vacuum expectation value of the field
commutator. It does not show the sharp discontinuity present when
one considers axially symmetric massless scalar fields in a
Minkowskian background and, hence, is a quantitative measure of
the spreading of the light cones expected in a quantized theory of
gravity as discussed at length in \cite{BarberoG:2004uc}.

\subsection{Asymptotic expansions in $\tau$}

In order to study the asymptotic behavior in $\tau$ for (\ref{f4})
it is convenient to consider two separate cases: Both $\rho_1$ and
$\rho_2$ different from zero or only one of them equal to zero
(the other must be different from zero because, otherwise, the
integral is badly divergent). In the first case the expansion can
be obtained by first writing (\ref{f4}) as an $h$-transform, use
the Mellin-Parseval formula to obtain a representation for it as a
complex contour integral, and split it (by displacing the
integration path close to the branch cuts of the integrand) into
several integrals that can be studied by standard Mellin-transform
techniques. The details of this procedure --that allows to obtain
the asymptotic behavior to any order-- can be found in
\cite{BarberoG:2004uc}. The result in the present case is
\begin{eqnarray*}
\frac{1}{2\pi\sqrt{\rho_1\rho_2}\log\tau}\left\{\exp\Big(\frac{\pi}{2}(\rho_1+\rho_2)
-i[\frac{\pi}{2}+\tau-(\rho_1+\rho_2)\log\tau]\Big)\Gamma[-i(\rho_1+\rho_2)]\right.\\
+\exp\Big(\!\!-\!\!\frac{\pi}{2}(\rho_1+\rho_2)
+i[\frac{\pi}{2}-\tau-(\rho_1+\rho_2)\log\tau]\Big)
\Gamma[i(\rho_1+\rho_2)]\hspace{-.1cm}\\
+\exp\Big(\frac{\pi}{2}(\rho_1-\rho_2)
-i[\tau+(\rho_2-\rho_1)\log\tau]\Big)\Gamma[i(\rho_2-\rho_1)]\hspace{.7cm}\\
\left.+\exp\Big(\frac{\pi}{2}(\rho_2-\rho_1)
-i[\tau+(\rho_1-\rho_2)\log\tau]\Big)
\Gamma[i(\rho_1-\rho_2)]\right\}+O(1/\log^2\tau).\hspace{-1.8cm}
\end{eqnarray*}
When either $\rho_1$ or $\rho_2$ are equal to zero we respectively
obtain
$$
\frac{1}{\sqrt{2\pi\rho_2\log\tau}}\left\{\exp\Big[\frac{\pi}{2}\rho_2
-i\big(\frac{\pi}{4}+\tau-\rho_2\log\tau\big)\Big]\Gamma(-i\rho_2)
+\exp\Big[\!-\!\frac{\pi}{2}\rho_2
+i\big(\frac{\pi}{4}-\tau-\rho_2\log\tau\big)\Big]\Gamma(i\rho_2)\right\}
+O(1/\log^{3/2}\tau)
$$
and
$$
\frac{1}{\sqrt{2\pi\rho_1\log\tau}}\left\{\exp\Big[\frac{\pi}{2}\rho_1
-i\big(\frac{\pi}{4}+\tau-\rho_1\log\tau\big)\Big]\Gamma(-i\rho_1)
+\exp\Big[\!-\!\frac{\pi}{2}\rho_1
+i\big(\frac{\pi}{4}-\tau-\rho_1\log\tau\big)\Big]\Gamma(i\rho_1)\right\}
+O(1/\log^{3/2}\tau).
$$
As we can see for fixed values of $\rho_1$ and $\rho_2$ the decay
when one of them is zero is slower (an inverse power of
$\sqrt{\log\tau}$) than the decay when both $\rho_1$ and $\rho_2$
are different from zero (an inverse power of $\log\tau$). We can
compare this result with the one corresponding to a massless,
axially symmetric free scalar field\footnote{Here we denote the
vacuum state of the system as $|0\rangle$.}
\begin{eqnarray*}
\langle0|\hat{\phi}(R_2,t_2)
\hat{\phi}(R_1,t_1)|0\rangle=\int_0^{\infty}\!\!\!
J_0(\rho_1q)J_0(\rho_2q)\exp(-i\tau
q)\,\mathrm{d}q\,.\label{f4-free}
\end{eqnarray*}
For values of $\tau$ satisfying $\tau>\rho_1+\rho_2$ this integral
is equal to\footnote{Complete elliptic integrals of the first,
second, and third kind are respectively defined as
$K(k)=\displaystyle\int_0^{\pi/2}\!\!\!\!\!\!\frac{\mathrm{d}\theta}{\sqrt{1-k^2\sin^2\theta}}$,
$E(k)=\displaystyle\int_0^{\pi/2}\!\!\!\!\sqrt{1-k^2\sin^2\theta}\,\mathrm{d}\theta$,
and
$\Pi(n|k)=\displaystyle\int_0^{\pi/2}\!\!\!\!\!\!\frac{\mathrm{d}\theta}{(1-n\sin^2\theta)\sqrt{1-k^2\sin^2\theta}}$.}
$$
-\frac{2i}{\pi\sqrt{\tau^2-(\rho_1-\rho_2)^2}}
\,K\!\left(\sqrt{\frac{4\rho_1\rho_2}{\tau^2-(\rho_1-\rho_2)^2}}\,\right)$$
and for large values of $\tau$ it behaves asymptotically as
$-i/\tau$, i.e. it falls off to zero much faster than (\ref{f4}).
As a consequence of this we interpret the very slow decay of the
two-point function for quantized Einstein-Rosen waves as an
enhanced probability amplitude to find quanta (either of the
gravitational or matter field) in the symmetry axis. This is a
gravitational effect as it is no present for a quantized, axially
symmetric, massless scalar field in a Minkowskian background.

\subsection{Asymptotic expansions for $\rho_1$, $\rho_2$,
and $\tau$ simultaneously large}

We discuss now the obtention of an asymptotic approximation that
is valid in a ``gravitational classical limit" corresponding to
taking $\rho_1$, $\rho_2$, and $\tau$ large while keeping their
relative values. This is equivalent to consider values for $R_1$,
$R_2$, and $t_2-t_1$ which are much larger than the ``Planck
scale" provided by $4G$; it is in this sense that we talk of a
classical limit here. To this end let us rewrite the integral in
(\ref{f4}) as
$$
\int_0^{\infty}\!\!\! J_0(\rho_1q)J_0(\rho_2q)\exp[-i\tau
(1-e^{-q})]\,\mathrm{d}q= \int_0^{\infty}\!\!\! J_0(\lambda
r_1q)J_0(\lambda r_2q)\exp[-i\lambda t (1-e^{-q})]\,\mathrm{d}q
$$
where $\rho_1=\lambda r_1$, $\rho_2=\lambda r_2$, and
$\tau=\lambda t$ with $r_1$, $r_2$, $t$ fixed, and $\lambda$ taken
as a new parameter that we will consider large (we will use it as
an asymptotic parameter $\lambda\rightarrow\infty$). The last
integral can be written as
\begin{eqnarray*}
-\frac{e^{-it\lambda}}{4\pi^2}\int_0^{\infty}\!\!\!
\mathrm{d}q\oint_{\gamma_1}\mathrm{d}z_1\oint_{\gamma_2}\mathrm{d}z_2
\frac{1}{z_1z_2}\,\exp\Big(
\lambda\Big[\frac{qr_1}{2}(z_1-\frac{1}{z_1})+\frac{qr_2}{2}(z_2-\frac{1}{z_2})
+ite^{-q}\Big]\Big)
\end{eqnarray*}
by using the well-know representation of the Bessel functions as
contour integrals (over contours $\gamma$ that enclose $z=0$)
$$
J_n(x)=\frac{1}{2\pi i}\oint_{\gamma}
\frac{\mathrm{d}z}{z^{n+1}}\exp
\Big[\displaystyle\frac{x}{2}\big(z-\frac{1}{z}\big)\Big].
$$
As discussed in \cite{BarberoG:2004uc} it is useful to chose the
contours $\gamma_{1,2}$ in the complex plane region satisfying
$\Re (z-1/z)\leq0,\,z\in\mathbb{C}$. A suitable asymptotic
expansion for it in $\lambda$ can then be obtained by following
the procedure outlined in \cite{BarberoG:2004uc}. We only quote
the result here after reabsorbing the parameter $\lambda$ (and,
hence, expressing the integral again in terms of $\rho_1$,
$\rho_2$, and $\tau$). To do this we need to define three
different regions that cover the ($\rho_1,\rho_2,\tau$)
space,\footnote{We restrict ourselves to positive values of $\tau$
--the extension to negative values is straightforward-- and avoid
the boundaries between these regions.} referred to in the
following as I, II, and III, and defined by the conditions:
$\tau\leq|\rho_2-\rho_1|$ for region I,
$|\rho_2-\rho_1|<\tau<\rho_1+\rho_2$ for region II, and
$\tau\geq\rho_1+\rho_2$ for region III. The asymptotic expansion
in each of them is given by the sum of a leading contribution
($\sim 1/\lambda$) and a first order correction behaving
asymptotically as $1/\lambda^{3/2}$ or $1/\lambda^2$. The leading
contribution in the different regions is
\begin{subequations}
\label{lambda_1}
\begin{eqnarray}
&&\mathbf{Region\, I}:\quad\quad
\frac{2}{\pi\sqrt{(\rho_1+\rho_2)^2-\tau^2}}\,
K\!\left(\sqrt{\frac{4\rho_1\rho_2}{(\rho_1+\rho_2)^2-\tau^2}}\,\right)\\
&&\mathbf{Region\, II}:\hspace{.5cm}
\frac{1}{\pi\sqrt{\rho_1\rho_2}}\left[\,
K\!\left(\sqrt{\frac{(\rho_1+\rho_2)^2-\tau^2}{4\rho_1\rho_2}}\,\right)-\!
iK\left(\sqrt{\frac{\tau^2-(\rho_2-\rho_1)^2}{4\rho_1\rho_2}}\,\right)\right]\\
&&\mathbf{Region\, III}:\hspace{.3cm}
\frac{-2i}{\pi\sqrt{\tau^2-(\rho_2-\rho_1)^2}}\,
K\!\left(\sqrt{\frac{4\rho_1\rho_2}{\tau^2-(\rho_2-\rho_1)^2}}\,\right)
\end{eqnarray}
\end{subequations}
and the first asymptotic correction
\begin{subequations}
\label{lambda_2}
\begin{eqnarray}
&&\hspace{-1cm}\mathbf{Region\, I}:\quad\quad
-\frac{i\tau}{2\pi}\left\{\frac{2\sqrt{(\rho_1+\rho_2)^2-\tau^2}
[\rho_1^4+\rho_2^4+2\rho_1^2\tau^2-3\tau^4+2\rho_2^2\tau^2-2\rho_1^2\rho_2^2]}
{(\rho_1+\rho_2-\tau)^2(\rho_1-\rho_2+\tau)^2(-\rho_1+\rho_2+\tau)^2(\rho_1+\rho_2+\tau)^2}\,
E\!\left(\sqrt{\frac{4\rho_1\rho_2}{(\rho_1+\rho_2)^2-\tau^2}}\,\right)\right.\nonumber\\
&&\hspace{2.4cm}\left.-\frac{2\tau^2}{\sqrt{(\rho_1+\rho_2)^2-\tau^2}
[\rho_2^4+(\tau^2-\rho_1^2)^2-2\rho_2^2(\rho_1^2+\tau^2)]}
\,K\!\left(\sqrt{\frac{4\rho_1\rho_2}{(\rho_1+\rho_2)^2-\tau^2}}\,\right)\right\}
                     \\
&&\hspace{-1cm}\mathbf{Region\,
II}:\hspace{.5cm}\frac{e^{i\left[\frac{\pi}{4}-\tau+|\rho_2-\rho_1|
(1+\log\frac{\tau}{|\rho_2-\rho_1|})\right]}}
{\sqrt{2\pi\rho_1\rho_2|\rho_1-\rho_2|}\log\frac{\tau}{|\rho_2-\rho_1|}}\\
&&\hspace{-1cm}\mathbf{Region\,
III}:\hspace{.3cm}\frac{1}{\sqrt{2\pi\rho_1\rho_2}}\left\{
\frac{e^{i\left[\frac{\pi}{4}-\tau+|\rho_2-\rho_1|(1
+\log\frac{\tau}{|\rho_2-\rho_1|})\right]}}
{\sqrt{|\rho_1-\rho_2|}\log\frac{\tau}{|\rho_2-\rho_1|}}+
\frac{e^{-i\left[\frac{\pi}{4}+\tau-(\rho_1+\rho_2)(1
+\log\frac{\tau}{\rho_1+\rho_2})\right]}}
{\sqrt{\rho_1+\rho_2}\log\frac{\tau}{\rho_1+\rho_2}} \right\}\,.
\end{eqnarray}
\end{subequations}
The imaginary part of these expressions gives the vacuum to vacuum
matrix elements of the field commutator already discussed in
\cite{BarberoG:2004uc}. Also it is important to point out that
(\ref{lambda_1}) is, precisely, the two-point function for an
axially symmetric massless scalar field evolving in a 2+1
dimensional Minkowskian background.  In fact, this is the leading
contribution to the two-point function --corresponding to an
ordinary quantum field theory for a massless scalar field in a
Minkowskian background-- whereas (\ref{lambda_2}) provide the
first asymptotic corrections.

It is also possible to obtain similar expansions when either
$\rho_1$ or $\rho_2$ are zero. In this case we only have a single
$J_0$ function in the integrand and the computations are greatly
simplified. The result is
\begin{equation}
\theta(\rho-\tau)\left[\frac{1}{\sqrt{\rho^2-\tau^2}}-
\frac{i\tau(\rho^2+2\tau^2)}{(\rho^2-\tau^2)^{5/2}}\right]
+\theta(\tau-\rho)
\left[\frac{-i}{\sqrt{\tau^2-\rho^2}}+
\frac{\exp(i[\rho\log\frac{\tau}{\rho}
-\tau+\rho])}{\rho\sqrt{\log\frac{\tau}{\rho}}}\right]
\end{equation}
where $\rho$ is the remaining non-zero radial parameter. Notice
that this cannot be obtained by simply putting $\rho_1=0$ or
$\rho_2=0$ in (\ref{lambda_1},\ref{lambda_2}).

\begin{figure}
\hspace{0cm}\includegraphics[width=17.5cm]{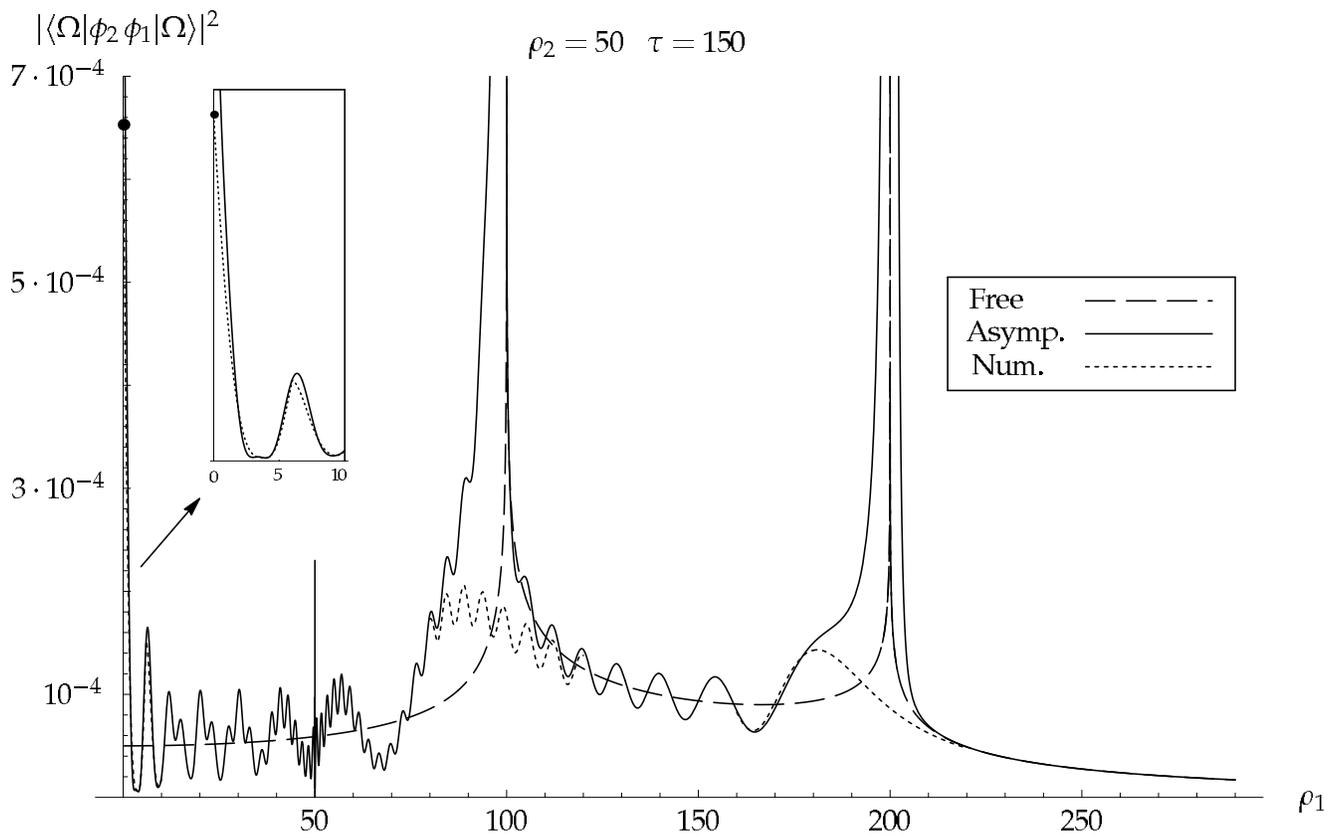}
\caption{This figure shows the absolute value squared for the
two-point function for fixed values of $\rho_2$ and $\tau$ in
terms of $\rho_1$. It shows the approximate probability of finding
a field quantum at a small volume centered around $\rho_1$ after a
certain time lapse $\tau$ if its position was $\rho_2$ at
$\tau=0$. Here radial distances and time are measured in units of
$4G$. We also compare the exact values obtained by numerical
computation (labelled as ``Num."), the approximation given by the
asymptotic expansions (labelled as ``Asymp."), and the one
corresponding to an axially symmetric massless scalar field
evolving in a Minkowskian background (labelled as ``Free"). The
most salient feature is the significant enhancement of the
probability for $\rho_1=0$. The dot on the vertical axis
corresponds to the value of the two-point function on the axis
$\rho_1=0$. The remarkable quality of the approximation provided
by the asymptotic expansions given in the paper can also be seen
except in the boundary between regions in the $\rho_1$, $\rho_2$,
and $\tau$ space, where the asymptotic expansions are divergent as
expected on general grounds. The inset shows in detail the
comparison of the exact values of the two-point function and the
asymptotic approximation given by (\ref{lambda_1},\ref{lambda_2}).
It is worthwhile to point out that even though one does not expect
the approximation to be valid for small values of $\rho_1$ or
$\rho_2$ its general behavior is well described by it. Notice also
the expected singularity at $\rho_1=\rho_2$.}
\label{fig1:twopoint}
\end{figure}

\begin{figure}
\hspace{0cm}\includegraphics[width=17.5cm]{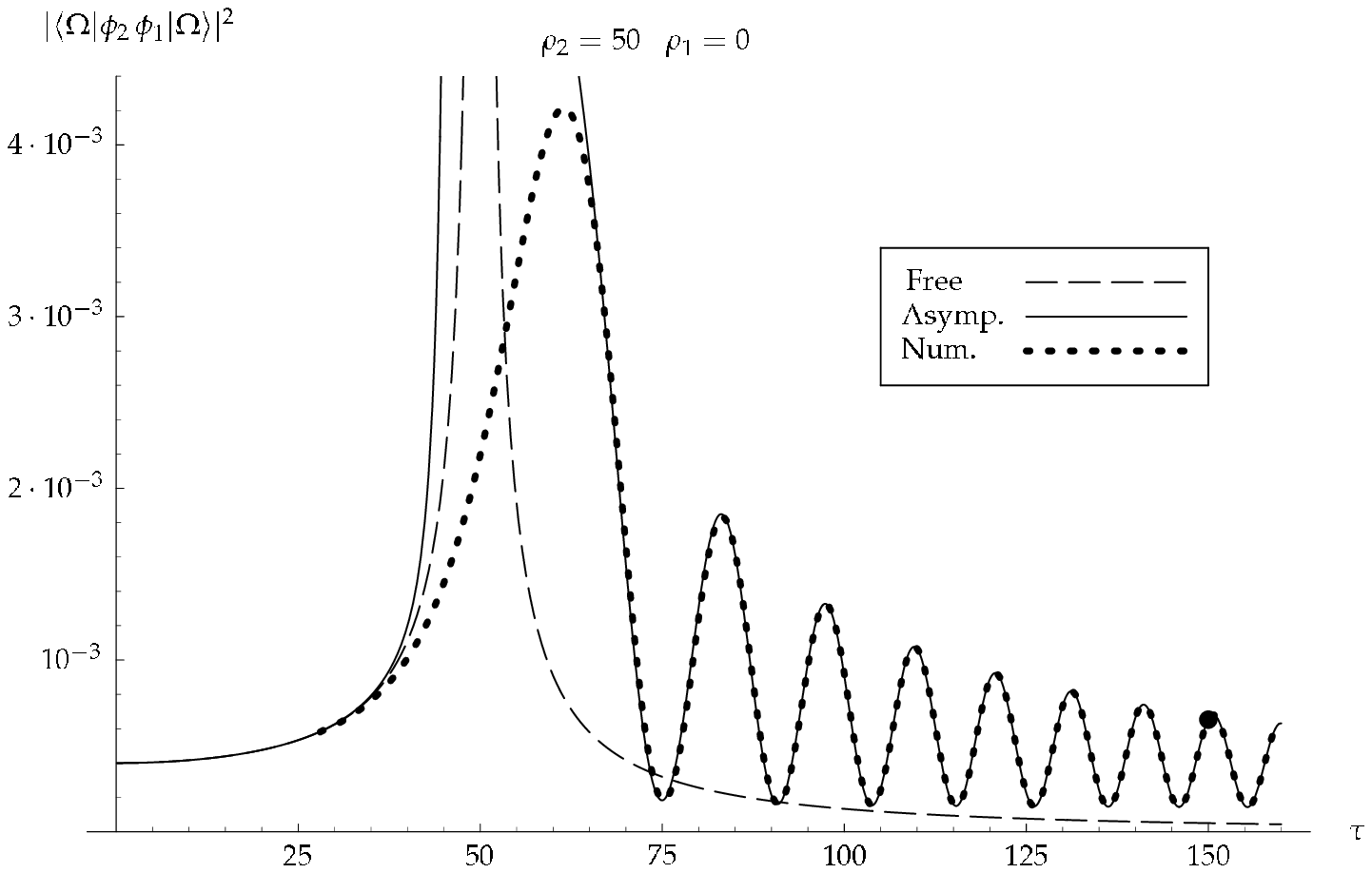}
\caption{This figure shows the absolute value squared for the
two-point function for a fixed value of $\rho_2$ and $\rho_1=0$.
The most interesting feature is the fact that the squared
amplitude does not behave in mean as the one corresponding to the
``free part" as was apparent in the previous picture. The dot at
$\tau=150$ corresponds to the value of the two-point function at
the symmetry axis shown in figure \ref{fig1:twopoint}. Notice the
different scale used in both plots.} \label{fig2:twopointaxis}
\end{figure}

Figures \ref{fig1:twopoint} and \ref{fig2:twopointaxis} show the
behavior of the two-point function. One can compare the exact
values obtained by numerical computation and the approximation
given by the asymptotic expansions. We also compare its value with
the one corresponding to an axially symmetric massless scalar
field evolving in a Minkowskian background. We emphasize again
here that the value of these asymptotic approximations relies on
the fact that they give the exact behavior of the two-point
functions in the relevant limits. The most important physical
information that can be gleaned from these plots is a significant
enhancement of the probability to find field quanta (either
gravitational or matter) in the vicinity of the symmetry axis
(defined by $\rho=0$) as compared with the result for an axially
symmetric massless scalar field in a Minkowskian background. This
is specially remarkable because far from the axis (both $\rho_1$
and $\rho_2$ large) the dominant contribution to the two-point
function is given by the one corresponding to the free massless
field. The asymptotic analysis for $\rho=0$ shows that the
symmetry axis is exceptional in the sense that one does not
recover the free-field result in the asymptotic limit. This was
already noticed in previous studies of the microcausality of the
system; in this context the interpretation of this phenomenon is
the enhanced probability mentioned above. Another feature that
stands out in the figures is the singularity at $\rho_1=\rho_2$.
This is expected on general grounds as a generic behavior in
quantum field theory. It can be removed by introducing suitable
regulators (see \cite{BarberoG.:2004uv}). Here we can, in
practice, identify and isolate the distributional behavior of the
relevant objects at these points so we will not introduce
regulators explicitly.

\bigskip

Other interesting two-point functions that we may study are mixed
ones involving both the gravitational and scalar fields. We then
consider matrix elements of the form $ \langle
f|\hat{\phi}_s(R_2;t_2,t_0)\hat{\phi}_g(R_1;t_1,t_0)|\Omega\rangle
$ for some state $|f\rangle$ different from the
vacuum.\footnote{Diagonal matrix elements of this type are always
zero --for both scalar and gravitational modes-- because they
involve products of an even number of creation operators with an
odd number of annihilation operators (or viceversa)  that act on
the vacuum to give zero. Also expectation values between the
vacuum and one particle states of either type are easily seen to
vanish.} The simplest non-zero matrix elements are obtained by
choosing
$$
|f\rangle=\int_0^{\infty}\!\!\mathrm{d}k_s\int_0^{\infty}\!\!\mathrm{d}k_g
f(k_s,k_g)A_s^{\dag}(k_s)A_g^{\dag}(k_g)|\Omega\rangle
$$
satisfying the normalization condition
$$
\int_0^{\infty}\!\!\int_0^{\infty}|f(k_s,k_g)|^2\,\mathrm{d}k_g\mathrm{d}k_s=1.
$$
They can be computed easily to give
\begin{eqnarray}
&&\hspace{-6mm}\langle
f|\hat{\phi}_s(R_2;t_2,t_0)\hat{\phi}_g(R_1;t_1,t_0)|\Omega\rangle\nonumber\\
&=& 4G\int_0^{\infty}\!\!\!\!\mathrm{d}k_s\!\!
\int_0^{\infty}\!\!\!\!\mathrm{d}k_gJ_0(R_2k_s)J_0(R_1k_g)\bar{f}(k_s,k_g)
\exp\bigg(i[(t_2-t_0)E(k_s)e^{-4Gk_g}+(t_1-t_0)E(k_g)]\bigg)\nonumber\\
&=&\frac{1}{4G}\int_0^{\infty}\!\!\!\!\mathrm{d}q_s\!\!
\int_0^{\infty}\!\!\!\!\mathrm{d}q_gJ_0(\rho_2q_s)J_0(\rho_1q_g)
\bar{f}(\frac{q_s}{4G},\frac{q_g}{4G})
\exp\bigg(i[\tau_{20}(1-e^{-q_s})e^{-q_g}+\tau_{10}(1-e^{-q_g})]\bigg)\label{mixedtwopoint}
\end{eqnarray}
where we have introduced adimensional variables and parameters as
above: $k_s=\frac{q_s}{4G}$, $k_g=\frac{q_g}{4G}$,
$\rho_1=\frac{R_1}{4G}$, $\rho_2=\frac{R_2}{4G}$,
$\tau_{10}=\frac{t_1-t_0}{4G}$, and
$\tau_{20}=\frac{t_2-t_0}{4G}$. It is also interesting to write
the matrix element of the commutator
\begin{eqnarray}
&&\hspace{-6mm}\langle
f|[\hat{\phi}_s(R_2;t_2,t_0),\hat{\phi}_g(R_1;t_1,t_0)]|\Omega\rangle\nonumber\\
&=&\hspace{-0mm}4G\int_0^{\infty}\!\!\!\!\!\!\mathrm{d}k_s\!\!
\int_0^{\infty}\!\!\!\!\!\!\mathrm{d}k_g\,J_0(R_2k_s)J_0(R_1k_g)\bar{f}(k_s,k_g)\bigg[
e^{i[(t_2-t_0)E(k_s)e^{-4Gk_g}+(t_1-t_0)E(k_g)]}-
e^{i[(t_2-t_0)E(k_s)+(t_1-t_0)E(k_g)e^{-4Gk_s}]}\bigg]\nonumber\\
&=&\hspace{-0mm}\frac{1}{4G}\int_0^{\infty}\!\!\!\!\!\!\mathrm{d}q_s\!\!
\int_0^{\infty}\!\!\!\!\!\!\mathrm{d}q_g\,J_0(\rho_2q_s)J_0(\rho_1q_g)
\bar{f}(\frac{q_s}{4G},\frac{q_g}{4G})\bigg[
e^{i[\tau_{20}(1-e^{-q_s})e^{-q_g}+\tau_{10}(1-e^{-q_g})]}-
e^{i[\tau_{20}(1-e^{-q_s})+\tau_{10}(1-e^{-q_g})e^{-q_s}]}\bigg]\label{mixedcomm}.
\end{eqnarray}

Some conclusions can be reached by considering some simple choices
for the function $f$ although a more detailed analysis would
require us to get suitable asymptotic expansions in terms of a
general $f$. To this end let us consider the normalized function
\begin{eqnarray*}
f(k_s,k_g)=\frac{1}{\hat{k}}\chi_{[k_{0s}-\hat{k}/2,k_{0s}+\hat{k}/2]}(k_s)
\chi_{[k_{0g}-\hat{k}/2,k_{0g}+\hat{k}/2]}(k_g)
\end{eqnarray*}
where $\chi_V$ is the characteristic function of the set $V$ and
$\hat{k}$ is a constant with dimensions of inverse length. For
values of $\hat{k}$ small enough that
$$
J_0(R_2k_s)J_0(R_1k_g)\exp\bigg(i[(t_2-t_0)E(k_s)e^{-4Gk_g}+(t_1-t_0)E(k_g)]\bigg)
$$
is essentially constant in the effective integration region the
value of the matrix elements (\ref{mixedtwopoint},\ref{mixedcomm})
are respectively
\begin{eqnarray*}
&&4G\hat{k}J_0(R_2k_{0s})J_0(R_1k_{0g})
e^{i[(t_2-t_0)E(k_{0s})e^{-4Gk_{0g}}+(t_1-t_0)E(k_{0g})]}\,,\\
&&4G\hat{k}J_0(R_2k_{0s})J_0(R_1k_{0g})\big\{
e^{i[(t_2-t_0)E(k_{0s})e^{-4Gk_{0g}}+(t_1-t_0)E(k_{0g})]}-
e^{i[(t_2-t_0)E(k_{0s})+(t_1-t_0)E(k_{0g})e^{-4Gk_{0s}}]}\big\}.
\end{eqnarray*}
and their squared amplitudes
\begin{eqnarray*}
&&16G^2\hat{k}^2J^2_0(R_2k_{0s})J^2_0(R_1k_{0g})\,,\\
&&8G^2\hat{k}^2J^2_0(R_2k_{0s})J^2_0(R_1k_{0g})\{1-\cos[4G(t_2-t_1)E(k_{0s})E(k_{0g})]\}.
\end{eqnarray*}
The first important point to notice here is the fact that these
quantities, and in particular the commutator, are generically
different from zero. This is an additional indication of the fact
that we are dealing with an interacting (i.e. non-free) theory. It
is also easy to see that the commutator is zero when $t_1=t_2$
because
$$
\exp\bigg(i(t_2-t_0)[E(k_s)e^{-4Gk_g}+E(k_g)]\bigg)=\exp\bigg(i(t_2-t_0)E(k_s+k_g)\bigg)
=\exp\bigg(i(t_2-t_0)[E(k_g)e^{-4Gk_s}+E(k_s)]\bigg).
$$
Finally it is remarkable that these functions do not display the
type of causal behavior that we have found for the commutator of
fields of the same type (by looking, for example, at the vacuum to
vacuum expectation value). Their magnitude is of order
$G^2\hat{k}^2$ so, at least in this approximation, are small in
the scale set by the natural length scales of the model.

\bigskip

\section{Newton-Wigner states: propagators}{\label{NWig}}

The main drawback of the two-point functions discussed in detail
in the previous section is the fact that their interpretation as
probability amplitudes is only approximate. The reason behind this
is the fact that the state vectors
$$
\hat{\phi}_{s,g}(R;t,t_0)|\Omega\rangle
$$
do not constitute an orthonormal set. This is at the root of the
well known problem of localization in relativistic quantum field
theory that has been solved in a more or less satisfactory way by
the so called Newton-Wigner states \cite{NW}. These constitute an
orthonormal basis of position eigenstates for a certain choice of
an inertial reference system in a Minkowskian spacetime. The most
important interpretive difficulty with them is the fact that they
cease to be localized under Lorentz boosts \cite{Teller:1995ti}
but the assumptions upon which their construction is based are so
natural that it is difficult to believe that a better solution to
this problem might exist.

The purpose of this section is to build localized states analogous
to the Newton-Wigner ones for our model, that we will label here
as $|R\rangle$. We have now a reduced spacetime symmetry group so
the problem of the behavior of localized states under spacetime
symmetry transformations is partially alleviated. The physical
interpretation of propagation amplitudes built with this type of
states is clear in the sense that they are now proper probability
amplitudes. Our point of view here is that the discussion of the
two-point function
$\langle\Omega|\hat{\phi}_s(R_2,t_2)\hat{\phi}_s(R_1,t_1)|\Omega\rangle$
together with the matrix elements $\langle R_2|\hat{U}(t_2,t_1)|
R_1\rangle$ can give us relevant and robust information about the
motion of field quanta in position space and provide meaningful
information about the transition between quantum and classical
geometry in quantum gravity. It should be pointed out here that
the availability of a position space orthonormal basis allows us
to define position state normalized wave functions
$$
|\Psi\rangle=\int_0^{\infty}\!\mathrm{d}R \,\Psi(R)|R\rangle,\quad
\langle R|\Psi\rangle=\Psi(R),\quad\int_0^{\infty}|\Psi(R)|^2\,
\mathrm{d}R=1
$$
and study their time evolution (in the Schr\"{odinger} picture)
given by $|\Psi(t)\rangle=\hat{U}(t,t_0)|\Psi(t_0)\rangle$.

Here we will follow a simple procedure inspired in \cite{NW}. Let
us write
$$\displaystyle|R\rangle=\int_0^{\infty}\!\mathrm{d}k\,f(k)J_0(kR)|k\rangle$$
where $|k\rangle:=|0\rangle^g\otimes|k\rangle^s$ are scalar matter
field quanta and we have made use of the fact that $J_0(kR)$ is a
solution of the radial part of the Schr\"{o}dinger equation for
states with zero angular momentum in 2 dimensions:
$$ \big[\partial_R^2+\frac{1}{R}\partial_R+k^2\big]J_0(kR)=0.
$$
Once we make this choice the function $f(k)$ is fixed by the
orthogonality condition $\langle R_2|R_1\rangle=\delta(R_2,R_1)$
which implies $|f(k)|^2=kR$, and hence
$f(k)=\sqrt{kR}e^{i\nu(k)}$. Without loss of generality we will
take $\nu(k)=0$. We finally get then
$$
|R\rangle=\int_0^{\infty}\!\mathrm{d}k\,(kR)^{1/2}J_0(kR)|k\rangle.
$$

In the following we will study matrix elements of the form
\begin{eqnarray}
\langle
R_2|\hat{U}(t_2,t_1)|R_1\rangle&\!\!=\!\!&\sqrt{R_1R_2}\int_0^{\infty}
\!kJ_0(kR_1)J_0(kR_2)\exp\big[-i(t_2-t_1)E(k)\big]\,\mathrm{d}k\nonumber\\
&=&\frac{\sqrt{\rho_1\rho_2}}{4G}\int_0^{\infty}\!q
J_0\big(\rho_1q\big)J_0\big(\rho_2q\big)
\exp[-i\tau(1-e^{-q})]\,\mathrm{d}q\label{NWintegral}\\
&=&\frac{e^{-i\tau}}{4G}\delta(\rho_1,\rho_2)+\frac{\sqrt{\rho_1\rho_2}}{4G}e^{-i\tau}
\int_0^{\infty}\!qJ_0\big(\rho_1q\big)J_0\big(\rho_2q\big)
\Big[\exp\big(i\tau e^{-q}\big)-1\Big]\,\mathrm{d}q.\nonumber
\end{eqnarray}
The last integral in the previous expression converges very
quickly as the integrand has an exponential decay. Notice also the
singularity at $\rho_1=\rho_2$ that was also present for the
two-point functions. As happened before we cannot give a closed
form expression for the integral (\ref{NWintegral}) although,
again, it can be computed numerically by essentially the same
methods used in the previous section. We can also obtain
asymptotic approximations of the types discussed above that help
us understand precisely the behavior in several important physical
regimes and in the ``classical limit".

An important question is the meaning of this probability
amplitude. In the case of the two-point function we highlighted
the interpretation of (\ref{radialpsi}) as the radial part of a
wave function with zero angular momentum for a free
two-dimensional particle. Now\footnote{Notice that $\langle
k|R\rangle$ is \textit{not} a solution to the Shr\"odinger
equation.} $|\langle k|R\rangle|^2=kRJ_0^2(kR)$ and the appearance
of the $R$ factor suggests that the correct interpretation for the
amplitudes given Newton-Wigner states is that they describe the
probability to find field quanta inside thin cylindrical shells at
a distance $R$. This means that we will have to introduce
appropriate factors of $R_1$ and $R_2$ to compare the two-point
functions of the previous section with $\langle
R_2|\hat{U}(t_2,t_1)|R_1\rangle$. Specifically we will study
\begin{equation}
\frac{(4G)^2}{\sqrt{R_1R_2}}\langle
R_2|\hat{U}(t_2,t_1)|R_1\rangle=\int_0^{\infty}\!q
J_0\big(\rho_1q\big)J_0\big(\rho_2q\big)
\exp\big[-i\tau(1-e^{-q})\big]\,\mathrm{d}q. \label{NWintegralmod}
\end{equation}
In the following we give the different asymptotic expansions for
this propagator in the same regimes described in the previous
section.

\subsection{Asymptotic expansions in $\rho_1$ or $\rho_2$}

For large values of $\rho_1$ and $\rho_2$ the asymptotic behaviors
of (\ref{NWintegralmod}) are respectively
$$
\frac{\tau}{\rho_1^3}\left\{i+\frac{9}{\rho_1^2}\left[
-\frac{i}{6}+\frac{i\rho_2^2}{4}+\frac{\tau}{2}+\frac{i\tau^2}{6}\right]\right\}+O(\rho_1^{-7}),\quad
\frac{\tau}{\rho_2^3}\left\{i+\frac{9}{\rho_2^2}\left[
-\frac{i}{6}+\frac{i\rho_1^2}{4}+\frac{\tau}{2}+\frac{i\tau^2}{6}\right]\right\}+O(\rho_2^{-7})
$$
obtained, again, by a straightforward application of Mellin
transform techniques. When this is compared to the asymptotic
behavior obtained for the two-point function we see that they
qualitatively agree for the imaginary part but they are quite
different for the absolute value or the real part. This is not
unexpected as the interpretation of the two-point function as a
probability amplitude is approximate.

\subsection{Asymptotic expansions in $\tau$}

As before it is convenient to study separately the case in which
both $\rho_1$ and $\rho_2$ are different from zero and the one in
which either $\rho_1$ or $\rho_2$ are zero. By using the same
methods as before we find

\begin{eqnarray*}
\frac{1}{2\pi\sqrt{\rho_1\rho_2}}\left\{\exp\Big(\frac{\pi}{2}(\rho_1+\rho_2)
-i[\frac{\pi}{2}+\tau-(\rho_1+\rho_2)\log\tau]\Big)
\Gamma[-i(\rho_1+\rho_2)]\right.\\
+\exp\Big(\!\!-\!\!\frac{\pi}{2}(\rho_1+\rho_2)
+i[\frac{\pi}{2}-\tau-(\rho_1+\rho_2)\log\tau]\Big)
\Gamma[i(\rho_1+\rho_2)]\hspace{-.1cm}\\
+\exp\Big(\frac{\pi}{2}(\rho_1-\rho_2)
-i[\tau+(\rho_2-\rho_1)\log\tau]\Big)\Gamma[i(\rho_2-\rho_1)]\hspace{.9cm}\\
\left.+\exp\Big(\frac{\pi}{2}(\rho_2-\rho_1)
-i[\tau+(\rho_1-\rho_2)\log\tau]\Big)\Gamma[i(\rho_1-\rho_2)]\right\}
+O(1/\log\tau).\hspace{-1.6cm}
\end{eqnarray*}
When either $\rho_1$ or $\rho_2$ are equal to zero we respectively
obtain
$$
\sqrt{\frac{\log
\tau}{2\pi\rho_2}}\left\{\exp\Big[\frac{\pi}{2}\rho_2
-i\big(\frac{\pi}{4}+\tau-\rho_2\log\tau\big)\Big]\Gamma(-i\rho_2)+\exp\Big[\!-\!\frac{\pi}{2}\rho_2
+i\big(\frac{\pi}{4}-\tau-\rho_2\log\tau\big)\Big]\Gamma(i\rho_2)\right\}+O(1/\log^{1/2}\tau)
$$
and
$$
\sqrt{\frac{\log
\tau}{2\pi\rho_1}}\left\{\exp\Big[\frac{\pi}{2}\rho_1
-i\big(\frac{\pi}{4}+\tau-\rho_1\log\tau\big)\Big]\Gamma(-i\rho_1)+\exp\Big[\!-\!\frac{\pi}{2}\rho_1
+i\big(\frac{\pi}{4}-\tau-\rho_1\log\tau\big)\Big]\Gamma(i\rho_1)\right\}+O(1/\log^{1/2}\tau).
$$

As we can see for fixed values of $\rho_1$ and $\rho_2$ when one
of them is zero the asymptotic behavior of (\ref{NWintegralmod})
for large values of $\tau$ consists of an oscillating function
times $\sqrt{\log\tau}$, also it must be noted that the
oscillating part is precisely the one that appeared in the
$\tau$-asymptotics of the two-point function. For $\rho_1$ and
$\rho_2$ both different from zero the asymptotic behavior is given
by factors that are purely oscillatory in $\tau$. Again these
coincide with those that appeared in the study of the two-point
function. We see now that the value in the axis grows (very
slowly) and it has a constant amplitude everywhere else. We see
again that there is an enhancement of the probability to find the
particle in the axis relative to the rest of the values of
$\rho_1$ and $\rho_2$. We can compare this result with the one
corresponding to the free axially symmetric scalar field in 2+1
dimensions given (for $\tau>\rho_1+\rho_2$) by
$$
\frac{2
\tau}{\pi[(\rho_1+\rho_2)^2-\tau^2]\sqrt{\tau^2-(\rho_1-\rho_2)^2}}
E\!\left(\sqrt{\frac{4\rho_1\rho_2}{\tau^2-(\rho_1-\rho_2)^2}}\,\right).
$$
As we can see for large values of $\tau$ it falls off to zero as
$-1/\tau^2$, much faster than the asymptotic expansions that we
have already found. Again we interpret this result as an enhanced
probability to find field quanta in the vicinity of the symmetry
axis. We see that although the analytic expressions that we have
obtained are different from the ones corresponding to the
two-point function the qualitative conclusions regarding the
behavior in the vicinity of the axis are the same. This strongly
suggests that we are seeing a genuine quantum gravitational
effect. The most significant difference between both types of
results is the fact that the probability at the axis decays very
slowly if one considers the two-point function whereas it slowly
grows if one uses the Newton-Wigner propagator. We do not perceive
a contradiction here because of the approximate interpretation of
the result for the two-point function and the necessity to look at
probabilities over spacetime regions. In fact we will look at this
again in the next section when we consider the evolution of actual
wave functions.

\subsection{Asymptotic expansions for $\rho_1$, $\rho_2$, and $\tau$ simultaneously large}

We discuss now the obtention of an asymptotic approximation that
is valid in a ``classical limit" corresponding to taking $\rho_1$,
$\rho_2$, and $\tau$ large while keeping their relative values as
we did for the two-point function. To this end let us rewrite the
integral in (\ref{NWintegralmod}) as
$$
\int_0^{\infty}\!\!\! q J_0(\rho_1q)J_0(\rho_2q)\exp[-i\tau
(1-e^{-q})]\,\mathrm{d}q= \int_0^{\infty}\!\!\! q J_0(\lambda
r_1q)J_0(\lambda r_2q)\exp[-i\lambda t (1-e^{-q})]\,\mathrm{d}q
$$
where $\rho_1=\lambda r_1$, $\rho_2=\lambda r_2$, and
$\tau=\lambda t$ with $r_1$, $r_2$, $t$ fixed, and $\lambda$ taken
as a new parameter that we will consider large (we will use it as
an asymptotic parameter $\lambda\rightarrow\infty$). The last
integral can be written as
\begin{eqnarray*}
-\frac{e^{-it\lambda}}{4\pi^2}\int_0^{\infty}\!\!\!
\mathrm{d}q\oint_{\gamma_1}\mathrm{d}z_1\oint_{\gamma_2}\mathrm{d}z_2
\frac{q}{z_1z_2}\,\exp\Big(
\lambda\Big[\frac{qr_1}{2}(z_1-\frac{1}{z_1})+\frac{qr_2}{2}(z_2-\frac{1}{z_2})
+ite^{-q}\Big]\Big)
\end{eqnarray*}
by using, again, the representation of the Bessel functions as
contour integrals. The leading contribution in the different
regions is
\begin{subequations}
\label{lambdaNW_1}
\begin{eqnarray}
&&\mathbf{Region\, I}:\quad\quad
\frac{2i\tau}{\pi[(\rho_1-\rho_2)^2-\tau^2]\sqrt{(\rho_1+\rho_2)^2-\tau^2}}\,
E\!\left(\sqrt{\frac{4\rho_1\rho_2}{(\rho_1+\rho_2)^2-\tau^2}}\,\right)\\
&&\mathbf{Region\, II}:\hspace{.5cm}
\frac{\tau}{\pi[(\rho_1-\rho_2)^2-\tau^2][(\rho_1+\rho_2)^2
-\tau^2]\sqrt{\rho_1\rho_2}}\times\nonumber\\
&&\hspace{2.5cm}\left\{\,
[(\rho_1+\rho_2)^2-\tau^2]K\!
\left(\sqrt{\frac{\tau^2-(\rho_1-\rho_2)^2}{4\rho_1\rho_2}}\,\right)-\!
4\rho_1\rho_2E\left(\sqrt{\frac{\tau^2-(\rho_2-\rho_1)^2}{4\rho_1\rho_2}}\,
\right)\right.\\
&&\hspace{2.7cm}\left.+i[(\rho_1-\rho_2)^2-\tau^2]
K\!\left(\sqrt{\frac{(\rho_1+\rho_2)^2-\tau^2}{4\rho_1\rho_2}}\right)+
4i\rho_1\rho_2E\left(\sqrt{\frac{(\rho_2
+\rho_1)^2-\tau^2}{4\rho_1\rho_2}}\,\right)\right\}\nonumber\\
&&\mathbf{Region\, III}:\quad\quad
\frac{2\tau}{\pi[(\rho_1+\rho_2)^2-\tau^2]\sqrt{\tau^2-(\rho_1-\rho_2)^2}}\,
E\!\left(\sqrt{\frac{4\rho_1\rho_2}{\tau^2-(\rho_1-\rho_2)^2}}\,\right)
\end{eqnarray}
\end{subequations}
and the first asymptotic correction
\begin{subequations}
\label{lambdaNW_2}
\begin{eqnarray}
&&\hspace{-1cm}\mathbf{Region\, I}:\quad\quad
\frac{\tau}{2}\frac{\partial^3}{\partial\tau^3}
\left[\frac{2}{\pi\sqrt{(\rho_1+\rho_2)^2-\tau^2}}\,
K\!\left(\sqrt{\frac{4\rho_1\rho_2}{(\rho_1+\rho_2)^2-\tau^2}}\,\right)\right]\\
&&\hspace{-1cm}\mathbf{Region\,
II}:\hspace{.5cm}\frac{e^{i\left[\frac{\pi}{4}-\tau+|\rho_2-\rho_1|
(1+\log\frac{\tau}{|\rho_2-\rho_1|})\right]}}
{\sqrt{2\pi\rho_1\rho_2|\rho_1-\rho_2|}}\\
&&\hspace{-1cm}\mathbf{Region\,
III}:\hspace{.3cm}\frac{1}{\sqrt{2\pi\rho_1\rho_2}}\left\{
\frac{e^{i\left[\frac{\pi}{4}-\tau+|\rho_2-\rho_1|
(1+\log\frac{\tau}{|\rho_2-\rho_1|})\right]}}
{\sqrt{|\rho_1-\rho_2|}}+
\frac{e^{-i\left[\frac{\pi}{4}+\tau-(\rho_1+\rho_2)
(1+\log\frac{\tau}{\rho_1+\rho_2})\right]}}
{\sqrt{\rho_1+\rho_2}} \right\}.
\end{eqnarray}
\end{subequations}
The previous expression for region I can be explicitly written in
terms of complete elliptic integrals of the first and second kind
with coefficients that are square roots of rational functions of
$\rho_1$, $\rho_2$, and $\tau$; as they are rather lengthy we do
not give them here. As before the leading contribution corresponds
to the Newton-Wigner propagator for a massless axially symmetric
scalar field evolving in a Minkowskian background. It is also
possible to give asymptotic expansions in the case when either
$\rho_1$ or $\rho_2$ are zero. They are
$$
\theta(\rho-\tau)\left[\frac{i\tau}{(\rho^2-\tau^2)^{3/2}}+
\frac{3\tau^2(3\rho^2+2\tau^2)}{2(\rho^2-\tau^2)^{7/2}}\right]+
\theta(\tau-\rho)\left[-\frac{\tau}{(\tau^2-\rho^2)^{3/2}}+
\frac{1}{\rho}e^{i(\rho-\tau+\rho\log\frac{\tau}{\rho})}
\sqrt{\log\frac{\tau}{\rho}}\,\right]
$$
where, as above, $\rho$ is the remaining non-zero radial
parameter. Notice that this last expansion cannot be obtained by
simply putting $\rho_1=0$ or $\rho_2=0$ in
(\ref{lambdaNW_1},\ref{lambdaNW_2}).

\begin{figure}
\hspace{0cm}\includegraphics[width=17.5cm]{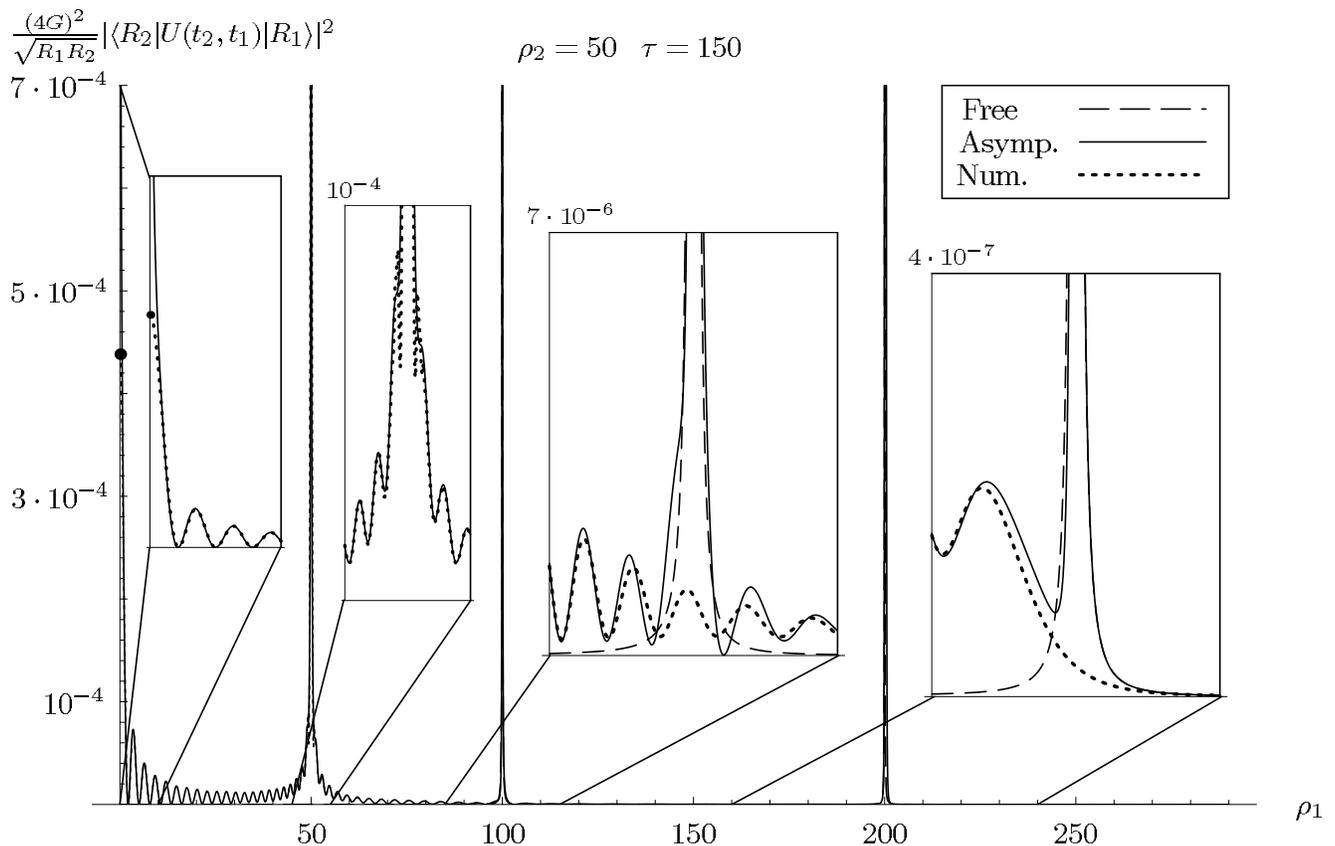}
\caption{This figure shows the square modulus of the Newton-Wigner
propagator (divided by factors of $R_{1,2}$ introduced in order to
compare it with the two-point function discussed in figure
\ref{fig1:twopoint}). We can see several interesting features: 1)
An enhanced amplitude at the axis similar to the one already seen
for the two-point function, 2) A large amplitude at
$\rho_1=\rho_2$, even when the delta function at this position is
subtracted; we interpret this a self gravity effect in a region of
hight matter density, 3) Beyond the position corresponding to
$\rho_1=200$ the amplitude decays very quickly; this marks the
position of the light cone. Notice that even though the amplitudes
in the ``free case" --corresponding to the propagation of a
massless axially symmetric scalar in a Minkowski background--
diverge in some regions in the $(\rho,\tau)$ plane  they remain
finite in our quantum gravity model.} \label{fig3:NWfueraeje}
\end{figure}
\begin{figure}
\hspace{0cm}\includegraphics[width=17.5cm]{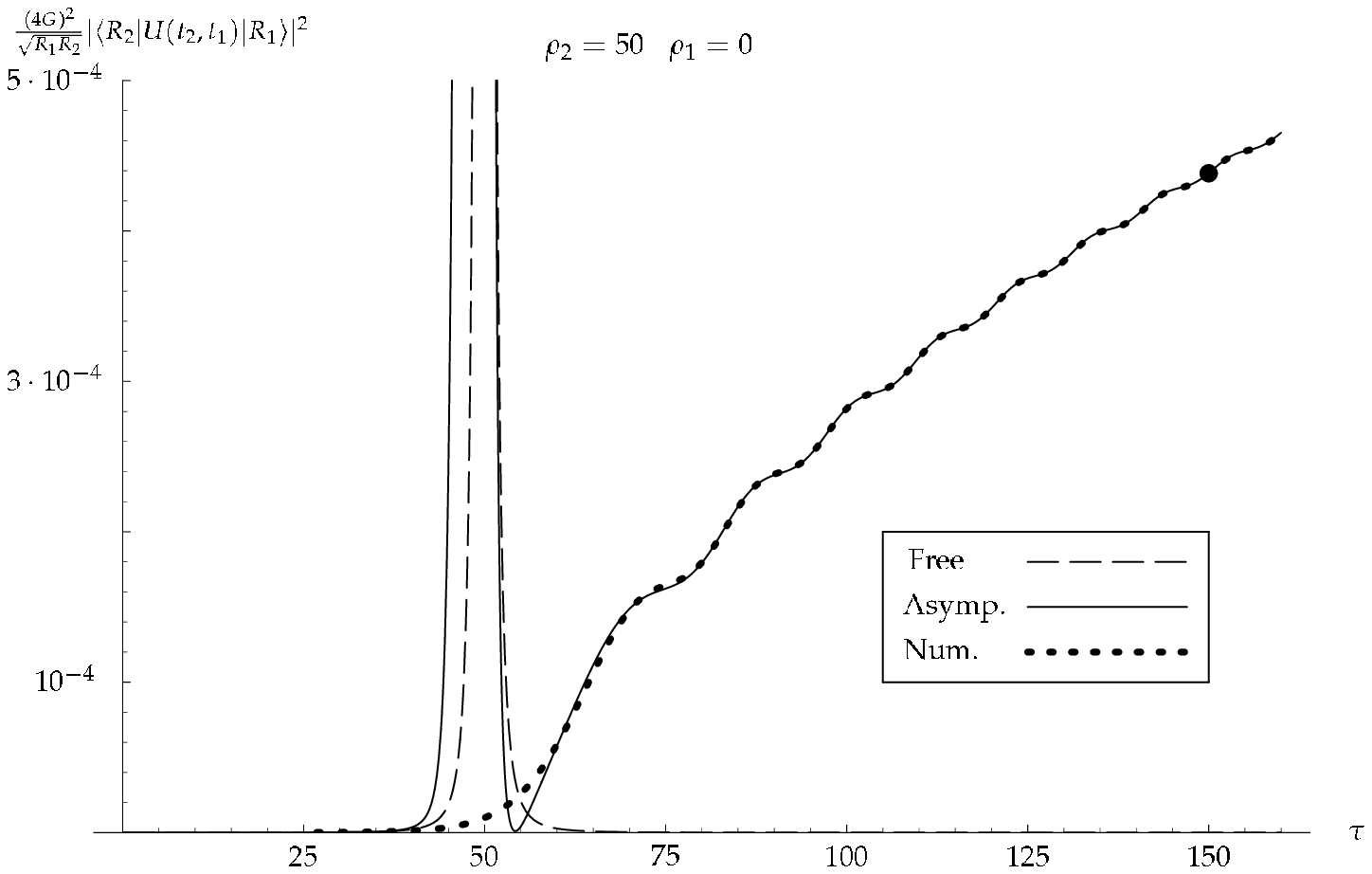}
\caption{This figure shows the square modulus of the Newton-Wigner
propagator (divided by factors of $R_{1,2}$) at the axis. The dot
corresponds to the value shown in figure \ref{fig3:NWfueraeje}.}
\label{fig4:NWpropeje}
\end{figure}

The most interesting feature of the propagator is its behavior in
the axis and at $\rho_1=\rho_2$. Let us consider first the
behavior at $\rho_1=\rho_2$. Here we find the singularity expected
on general grounds due to the orthonormality property of the
Newton-Wigner vectors. This can be identified as the delta
function appearing in (\ref{NWintegral}). In addition to this we
see a clear tendency of the probability amplitude to remain
concentrated around the region $\rho_1=\rho_2$. We interpret this
as an effect of self gravity that tends to favor the concentration
of matter. The spreading of the amplitude, on the other hand, can
be interpreted as quantum mechanical diffusion similar, but less
extreme, than the familiar one for particles in ordinary quantum
mechanics. At the axis $\rho_1=0$ (or $\rho_2=0$) we see that once
the amplitude grows as a consequence of the gravitational collapse
of the initial matter distribution there is a tendency to have a
large probability to find field quanta at that position (in fact
the amplitude grows!). To really assess how the probability of
finding particles near the symmetry axis evolves we will consider
in the next section the evolution of proper normalizable wave
functions and confirm, indeed that the probability is enhanced but
decays very slowly in time. These effects bear some resemblance
with what one expects to find in the study of gravitational
collapse and black hole evaporation.\footnote{The main difference
is the apparent absence of a horizon or something behaving, at
least in an approximate way, as one.} It is important to compare
the results of this section with the ones derived in the study of
the two-point function and the free axially symmetric massless
scalar field. Regarding the first it should be pointed out that
the qualitative agreement between the two pictures is very good.
We see in both cases the enhancement of the probability at the
axis, the singular structure in the vicinity of $\rho_1=\rho_2$,
and the motion along null radial geodesics (manifest itself as a
significant probability to find particles at the ``classical"
light cone). When the results are compared to the ``free" massless
case the gravitational phenomena that we have interpreted as an
increased probability due to self-gravity are conspicuously absent
but those related to the causal structure and microcausality of
the system (i.e. enhanced probabilities on the light cones) are
still present.

\bigskip

\section{Time evolution of a (radial) position state wave
function}{\label{Wave}}

\bigskip

For superpositions of one particle states the time evolution of
the wave function $\Psi(R,t)=\langle R|\Psi(t)\rangle$ is given by
$$
\Psi(R,t)=\langle R
|\hat{U}(t,t_0)|\Psi(t_0)\rangle=\int_0^\infty\!\!\!
\int_0^\infty\!\!\! k\sqrt{R
\tilde{R}}\,J_0(kR)J_0(k\tilde{R})\exp[-i(t-t_0)E(k)]\,\Psi(\tilde{R},t_0)\,\mathrm{d}
\tilde{R}\,\mathrm{d}k\,.
$$
In the following we want to study the evolution of one particle
wave functions of this type. This will be, in fact, the quantum
test particles that we will use  to describe the quantum geometry
of our spacetime model. In principle we can make any choice of
initial wave function $\Psi(R,t_0)$. However, in order to obtain
closed expressions for the wave function (or at least as simple as
possible) we will concentrate on a specific choice that satisfies
several reasonable requirements: The possibility of having some
control on the position of the peak of the probability
distribution, the possibility of controlling the width of the wave
packet, and the possibility of performing (some) integrations to
get a manageable closed form for it. A (normalized) function
satisfying these conditions at $t_0$ is
$$\Psi(R,t_0)=\sqrt{\frac{2R}{r_2^2-r_1^2}}\,\chi_{[r_1,r_2]}(R)
\quad\mathrm{  (with}\, r_2>r_1\mathrm{)}$$
that gives
\begin{eqnarray*}
\Psi(R,t)=\sqrt{\frac{2R}{r_2^2-r_1^2}}\int_0^{\infty}\!\!\!
\!J_0(kR)[r_2J_1(kr_2)-r_1J_1(kr_1)]
\exp[-i(t-t_0)E(k)]\,\mathrm{d}k.
\end{eqnarray*}
It is convenient to rewrite this in terms of adimensional objects
by introducing the following redefinitions  $\rho=\frac{R}{4G}$,
$\sigma_2=\frac{r_2}{4G}$, $\sigma_1=\frac{r_1}{4G}$,
$\tau=\frac{t-t_0}{4G}$ and the change of integration variable
$q=4Gk$. Writing $\psi(\rho,\tau):=\Psi(4G\rho,4G\tau+t_0)$ we
finally get
\begin{equation}
\psi(\rho,\tau)=\sqrt{\frac{2\rho}{4G(\sigma_2^2-\sigma_1^2)}}\int_0^{\infty}\!\!\!\!
J_0(q\rho)[\sigma_2J_1(q\sigma_2)-\sigma_1J_1(q\sigma_1)]
\exp[-i\tau(1- e^{-q})]\,\mathrm{d}q \label{psi}
\end{equation}
that satisfies, for all $\tau$, the normalization condition
$$
4G\int_0^{\infty}|\psi(\rho,\tau)|^2\mathrm{d}\rho=1.
$$
It is straightforward to see that when $\tau=0$ we recover the
initial wave function at $t=t_0$. Our goal now is to extract
information about the evolution of this wave function. In
particular we are interested in the behavior near the symmetry
axis to see if the enhanced probability suggested by the analysis
of the two-point functions and the Newton-Wigner propagator is
present. We also want to find out if the evolution of this wave
function somehow defines a classical trajectory in the spacetime
that can be used to define in an approximate way a physical notion
of geodesic. Finally we want to compare the result with the one
obtained for the axially symmetric massless scalar that we are
using to disentangle quantum gravitational effects from more
prosaic behaviors. As in the previous sections we will use
asymptotic approximations in different regimes to extract the
analytic behavior of $\psi(\rho,\tau)$.

\subsection{Asymptotic expansion in $\rho$}

For large values of $\rho$ the asymptotic behavior of (\ref{psi})
is
$$
\frac{1}{2}\sqrt{\frac{\sigma_2^2-\sigma_1^2}{4G}}
\frac{\tau}{\rho^{3/2}} \left\{
i+\frac{9}{2\rho^2}\left[-\frac{i}{3}+\tau+\frac{i\tau^2}{3}
+\frac{9}{4}(\sigma_1^2+\sigma_2^2)\right]
\right\}+O(\rho^{-9/2})
$$
obtained, again, by a straightforward application of Mellin
transform techniques. When this is compared to the asymptotic
behavior obtained for the evolution of the same wave function for
the case of an axially symmetric massless scalar in a Minkowskian
background
$$
\frac{1}{2}\sqrt{\frac{\sigma_2^2-\sigma_1^2}{4G}}
\frac{\tau}{\rho^{3/2}} \left\{
i+\frac{9}{2\rho^2}\left[\frac{i\tau^2}{3}+\frac{9}{4}(\sigma_1^2+\sigma_2^2)\right]
\right\}+O(\rho^{-9/2})
$$
we see that the leading behavior far from the axis (large $\rho$
and fixed $\tau$) is the same in both cases.

\subsection{Asymptotic expansions in $\tau$}

In principle we only have to consider the situation in which $\rho
\neq 0$ because $\psi(0,\tau)=0$. The asymptotic behavior in
$\tau$ for the integral
\begin{equation}
\int_0^{\infty}J_0(\rho q)J_1(\sigma q)\exp[-i\tau(1-e^{-q})]\,
\mathrm{d}q \label{int_psi}
\end{equation}
is obtained by the same methods used in previous sections and is
\begin{eqnarray*}
&&S(\rho,\sigma,\tau)=\frac{1}{2\pi\sqrt{\rho\sigma}\log\tau}
\left\{-\exp\Big(\frac{\pi}{2}(\rho+\sigma)
+i[-\tau+(\rho+\sigma)\log\tau]\Big)\Gamma[-i(\rho+\sigma)]\right.\\
&&\hspace{4cm}-\exp\Big(\!\!-\!\!\frac{\pi}{2}(\rho+\sigma)
-i[\frac{\pi}{2}\tau+(\rho+\sigma)\log\tau]\Big)\Gamma[i(\rho+\sigma)]\hspace{-.1cm}\\
&&\hspace{4cm}+\exp\Big(\frac{\pi}{2}(\rho-\sigma)
+i[-\tau+(\rho-\sigma)\log\tau+\frac{\pi}{2}]\Big)
\Gamma[i(\sigma-\rho)]\hspace{.7cm}\\
&&\hspace{4cm}\left.+\exp\Big(\frac{\pi}{2}(\sigma-\rho)
-i[\tau+(\rho-\sigma)\log\tau+\frac{\pi}{2}]\Big)\Gamma[i(\rho-\sigma)]
\right\}+O(1/\log^2\tau),\hspace{-1.8cm}
\end{eqnarray*}
so that we have that the asymptotic behavior for $\psi(\rho,\tau)$
is given by
$$
\sqrt{\frac{2\rho}{4G(\sigma_2^2-\sigma_1^2)}}\left[\sigma_2
S(\rho,\sigma_2,\tau)-\sigma_1S(\rho,\sigma_1,\tau)\right].
$$
This displays the slow decay in time that is characteristic of the
system. As before one can compare it with the one corresponding to
the free, massless, axially symmetric scalar field that is given
for large values of $\tau$ by
\begin{eqnarray*}
&&\frac{2}{\pi}\sqrt{\frac{2\rho}{4G(\sigma_2^2-\sigma_1^2)}}
\left[\,\frac{(\rho +\sigma_2 -\tau ) \Pi \left(\frac{2 \rho
}{\rho -\sigma_2 +\tau }\Big|\sqrt{\frac{4 \rho \sigma_2}{\tau
^2-(\rho -\sigma_2)^2}}\right)-\sigma_2 K\left(\sqrt{\frac{4 \rho
\sigma_2}{\tau ^2-(\sigma_2-\rho)^2}}\right)}{\sqrt{\tau
^2-(\sigma_2-\rho)^2}}\right.\hspace{1cm}\\
&&\hspace{2.7cm}\left.-\frac{(\rho +\sigma_1 -\tau ) \Pi
\left(\frac{2 \rho }{\rho -\sigma_1 +\tau }\Big|\sqrt{\frac{4 \rho
\sigma_1}{\tau ^2-(\rho -\sigma_1)^2}}\right)-\sigma_1
K\left(\sqrt{\frac{4 \rho \sigma_1}{\tau
^2-(\sigma_1-\rho)^2}}\right)}{\sqrt{\tau
^2-(\sigma_1-\rho)^2}}\,\right]
\end{eqnarray*}
and decays to zero as
$$\displaystyle{\frac{\sigma_1^2-\sigma_2^2}{2\tau^2}\sqrt{\frac{2\rho}{4G(\sigma_2^2-\sigma_1^2)}}}.$$

This behavior in $\tau$ means that if the evolution of the initial
wave function is such that at some instant of time $\tau$ the
probability of finding the particle in the vicinity of the
symmetry axis builds up, it will remain high for a large interval
in $\tau$ as the asymptotic behavior obtained above shows. In fact
this is what happens in this case as can be seen in figure
\ref{fig5:onda}. Finally it is interesting to notice that the
fall-off in the $\tau$ direction is much faster than the one given
by the $\tau$ asymptotic expansion for our system.

\subsection{Asymptotic expansions for $\rho$, $\sigma$, and $\tau$ simultaneously large}

We discuss now the obtention of an asymptotic approximation that
is valid in a ``classical limit" corresponding to taking $\rho$
and $\tau$ large while keeping their relative values as we did in
previous sections. We will also take $\sigma_1$ and $\sigma_2$,
that define the support of the wave function at the initial time,
large in comparison with the length scale $4G$. This will allow us
to use the same type of asymptotic expansion that we have used in
previous sections. The procedure should be clear by now so we skip
the details here, we just use the contour integral representation
introduced above for the Bessel functions and write the integrals
in (\ref{psi}) in terms of them. The asymptotic expansion for
(\ref{int_psi}) is then obtained as the sum of a boundary term
contribution that coincides with
\begin{eqnarray*}
\int_0^{\infty}J_0(\rho q)J_1(\sigma q)\exp(-i\tau q)\,
\mathrm{d}q
\end{eqnarray*}
plus some extra contributions. The boundary term is
\begin{subequations}
\label{lambda_psi}
\begin{eqnarray}
&&|\rho-\sigma|>\tau:\hspace{.3cm}
\frac{\mathrm{sgn}(\sigma-\rho)}{\sigma }+\frac{2 i \left[\sigma
K\left(\sqrt{\frac{4 \rho  \sigma }{(\rho +\sigma )^2-\tau
^2}}\right)+(\rho -\sigma -\tau ) \Pi \left(\frac{2 \rho }{\rho
+\sigma +\tau }\Big|\sqrt{\frac{4 \rho \sigma
   }{(\rho +\sigma )^2-\tau ^2}}\right)\right]}{\pi  \sigma
   \sqrt{(\rho +\sigma )^2-\tau ^2}}\\
&&\hspace{-1.3cm}|\rho-\sigma|<\tau<\rho+\sigma:\quad
\frac{1}{\sigma }+\frac{i (\rho +\tau ) \left[i
K\left(\sqrt{\frac{\tau ^2-(\rho -\sigma )^2}{4 \rho \sigma
}}\right)-K\left(\sqrt{\frac{(\rho +\sigma )^2-\tau ^2}{4 \rho
\sigma }}\right)\right]}{\pi  \sigma
   \sqrt{\rho  \sigma }}\nonumber\\
   &&\hspace{2.2cm}+\frac{i \left[(\rho +\sigma +\tau )
   K\left(\sqrt{\frac{(\rho +\sigma )^2-\tau ^2}{4 \rho  \sigma }}\right)
   +(\rho -\sigma -\tau ) \Pi \left(\frac{\rho +\sigma
   -\tau }{2 \sigma }\Big|\sqrt{\frac{(\rho +\sigma )^2-\tau ^2}
   {4 \rho  \sigma }}\right)\right]}{\pi  \sigma  \sqrt{\rho  \sigma }}\\
&&\hspace{2.2cm}-\frac{(\rho +\sigma -\tau ) \Pi \left(\frac{\rho
-\sigma
   +\tau }{2 \rho }\Big|\sqrt{\frac{\tau ^2-(\rho -\sigma )^2}{4 \rho  \sigma }}\right)-
   2 \rho  K\left(\sqrt{\frac{\tau ^2-(\rho -\sigma )^2}{4 \rho  \sigma }}\right)}
   {\pi  \sigma  \sqrt{\rho\sigma }}\nonumber\\
&&\hspace{.2cm}\rho+\sigma<\tau:\hspace{.3cm}\frac{1}{\sigma
}+\frac{2 \left[(\rho +\sigma -\tau ) \Pi \left(\frac{2 \rho
}{\rho -\sigma +\tau }\Big|\sqrt{\frac{4 \rho \sigma }{\tau
^2-(\rho -\sigma )^2}}\right)-\sigma K\left(\sqrt{\frac{4 \rho
\sigma}{\tau ^2-(\sigma -\rho )^2}}\right)\right]}{\pi  \sigma
\sqrt{\tau ^2-(\sigma -\rho )^2}}
\end{eqnarray}
\end{subequations}
and the extra contribution is given by
\begin{eqnarray*}
&&\theta(\tau-\sigma-\rho)
\frac{\exp\left(i[-\tau+(\rho+\sigma)(1+\log\frac{\tau}{\rho+\sigma})
-\frac{3\pi}{4}]\right)}{\sqrt{2\pi\sigma\rho(\sigma+\rho)}\log\frac{\tau}{\rho+\sigma}}\\
&+&\theta(\tau-\sigma+\rho)\theta(\sigma-\rho)
\frac{\exp\left(i[-\tau+(\sigma-\rho)(1+\log\frac{\tau}{\sigma-\rho})
-\frac{\pi}{4}]\right)}{\sqrt{2\pi\sigma\rho(\sigma-\rho)}\log\frac{\tau}{\sigma-\rho}}
\\&+&
\theta(\tau+\sigma-\rho)\theta(\rho-\sigma)
\frac{\exp\left(i[-\tau+(\rho-\sigma)(1+\log\frac{\tau}{\rho-\sigma})
+\frac{3\pi}{4}]\right)}{\sqrt{2\pi\sigma\rho(\rho-\sigma)}\log\frac{\tau}{\rho-\sigma}}.
\end{eqnarray*}
Introducing these expressions in (\ref{psi}) we finally obtain the
desired asymptotic approximation for the wave function in the
limit when all the lengths are significantly larger than $4G$.

\begin{figure}
\hspace{0cm}\includegraphics[width=17.5cm]{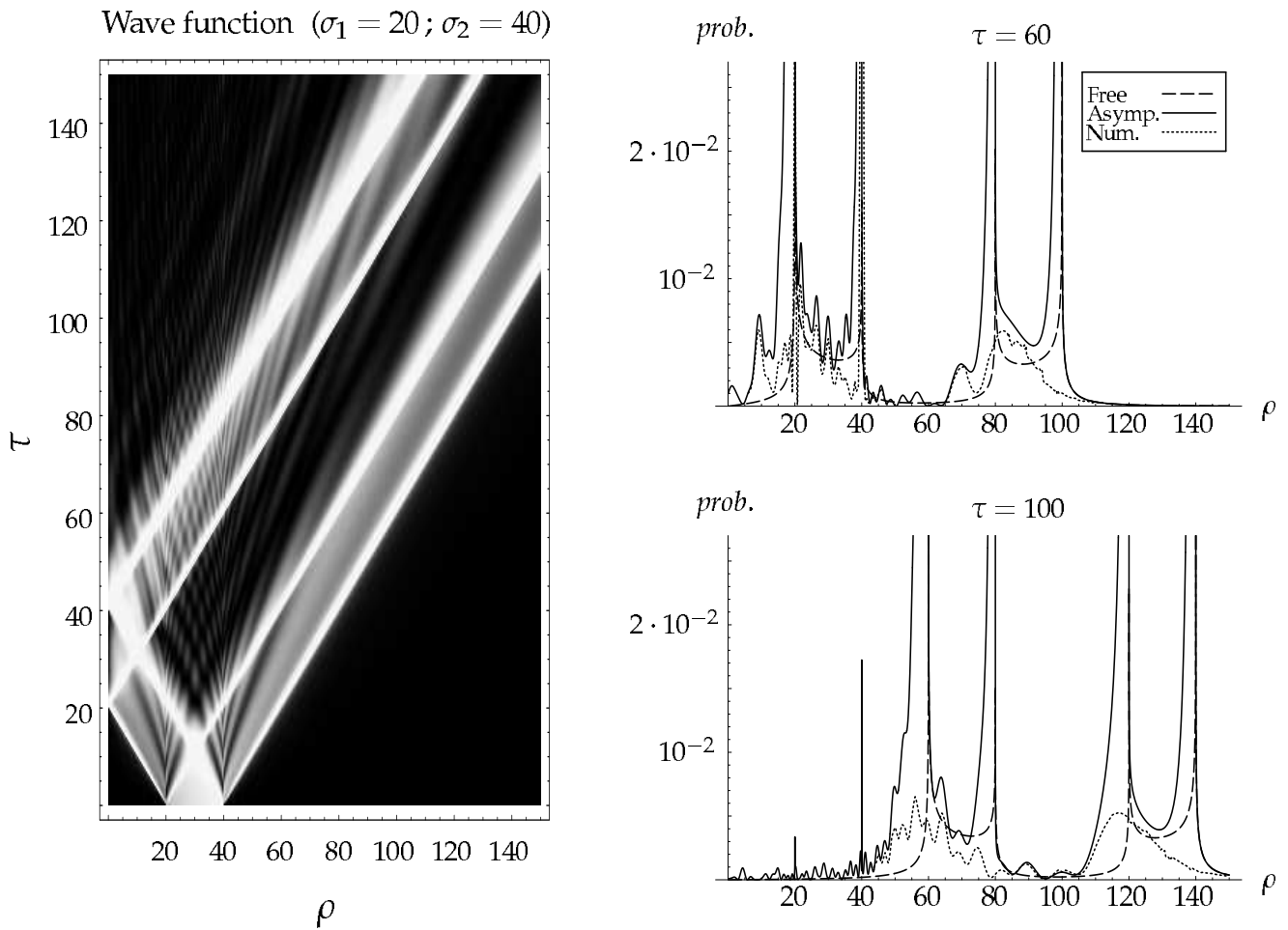}
\caption{Probability density for a wave function with a wide
support at the initial instant of time. The density plot is made
with the asymptotic approximation discussed in the text to avoid
long and slow numerical integrations. Saturated (white) areas
correspond to divergencies of this asymptotic expansion that are
not present in the true wave function. This can be seen in the
sections plotted in the right hand side of the figure. Notice that
in this case the two null curves signalling the light cone are
well defined --certainly better than in the narrow case shown in
figure \ref{fig5:onda}-- and the probability inside the light cone
or at the axis becomes very low as $\tau$ grows.}
\label{fig6:onda}
\end{figure}

\begin{figure}
\hspace{0cm}\includegraphics[width=17.5cm]{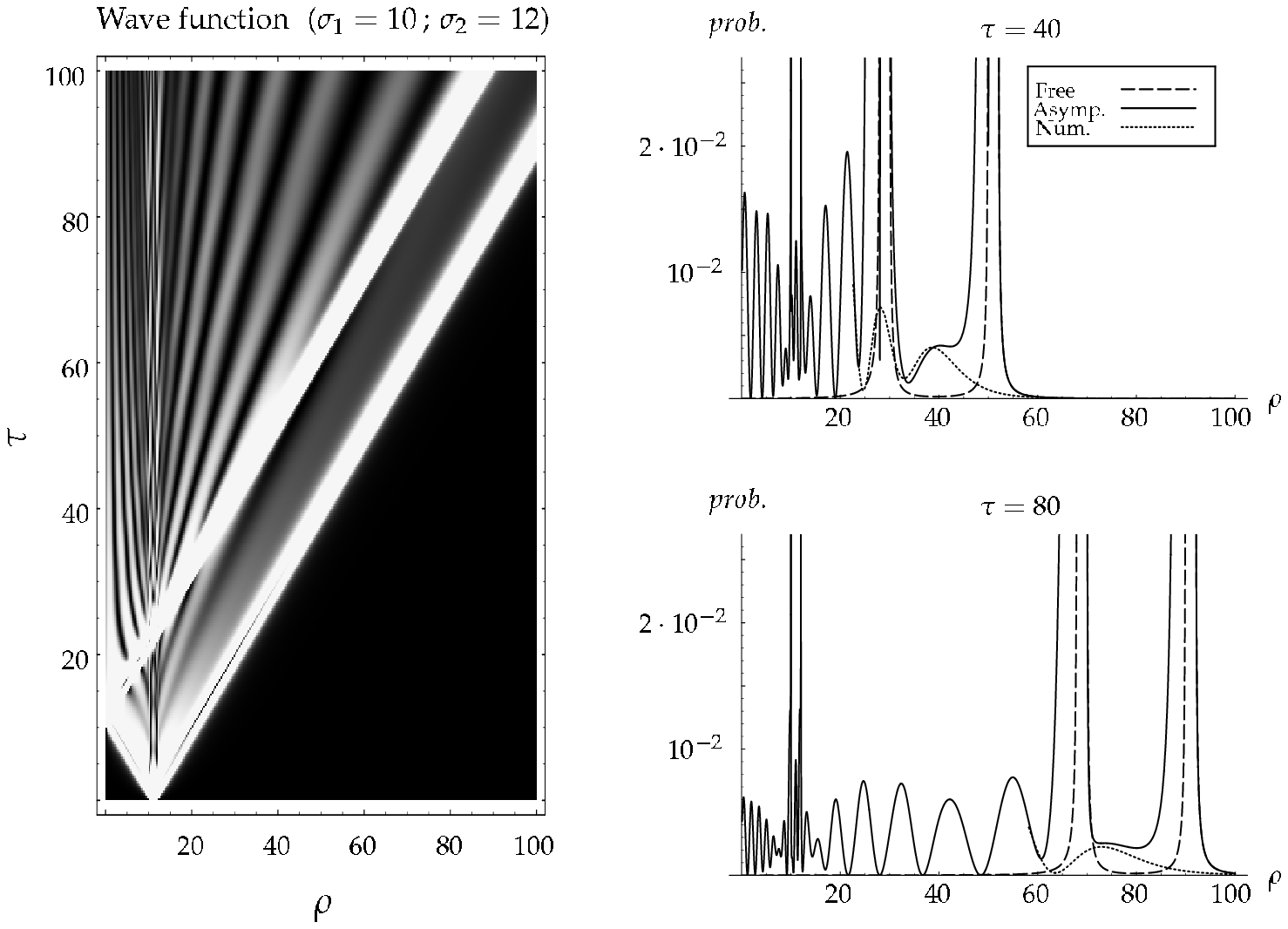}
\caption{Probability density for an initial wave function with a
narrow support at the initial instant of time. Notice the enhanced
probability close to the axis and the position of the light cone
structure that appears.} \label{fig5:onda}
\end{figure}

These asymptotic expansions allow us to explore different
possibilities  as far as the width and the position of the support
of the wave function at $t=t_0$ is concerned. We may consider the
case in which the support  --in the scale defined by $4G$-- is
wide or narrow.\footnote{The Newton-Wigner propagator $\langle
R_2|\hat{U}(t_2,t_1)|R_1\rangle$ corresponds to the limit when the
support is infinitely narrow.} In the first case the wave function
evolves in a way that closely resembles the propagation of an
initial wave function of this type for a free axially symmetric
field in a Minkowski background. One can easily see that most of
the probability amplitude for large values of $\rho$ is
concentrated along the lines $\tau=\rho_0+\rho$ and
$\tau=\rho-\rho_0$ in the $(\rho,\tau)$ plane. These two lines
define trajectories that can be interpreted as null geodesics of
an emergent spacetime metric. Notice that they are defined with a
resolution of the order of the width of the initial support of the
wave function (see figure \ref{fig6:onda}). The other case defines
a situation when the matter density is in some sense high and then
displays a behavior that can be interpreted as due to self gravity
effects (see figure \ref{fig5:onda}). Also in this case, specially
when the initial support is close to the axis, there is a build up
of the probability amplitude at $\rho=0$ that decays subsequently
in the very slow fashion characteristic of the model. This means
that the probability to find the particle in the vicinity of the
axis remains high for a long time and, as a consequence, the
probability of finding it on the ``light cone" is much lower. This
is shown in figure \ref{fig5:onda}.

As we have already discussed in the case of the two-point function
and the Newton-Wigner propagator there is significant enhancement
of the probability to find field quanta close to the axis. This
can easily be seen by comparing this wave function to the one
corresponding to quanta of an axially symmetric massless scalar
field in a 2+1 dimensional Minkowskian background. As the
asymptotic behavior of both shows (and the figures clearly
display) the probability near the axis remains significantly
higher in the gravitational case if the support of the initial
wave function is narrow. Another interesting feature that can be
seen is the persistence of a footprint of the wave function in the
range of $\rho$ where the support of the initial wave function is.
This can be understood by realizing that it is possible to write
the Newton-Wigner propagator (\ref{NWintegral}) as the sum of a
delta function, multiplied by a time-dependent phase, and a
rapidly convergent integral.

\section{Conclusions and comments}{\label{Conclu}}

In the first part of the paper we have discussed in some detail
the exact non-perturbative quantization of Einstein-Rosen waves
coupled to a massless, axially symmetric scalar field. This
provides us with an interesting system where gravity is coupled to
a matter field that can be exactly quantized and retains some of
the features of full general relativity. In particular, and even
though we have performed a symmetry reduction, we still have and
infinite number of local degrees of freedom both in the
gravitational and matter sector. We also have (some)
diffeomorphism invariance that has been taken care of through
gauge fixing in the Hamiltonian formalism.

One of the uses of this toy model is to explore the possibility of
getting relevant information on the nature of quantum space time
by using the probe provided by the matter field. This is
particularly useful in the setting of Einstein-Rosen waves owing
to the fact that the object that encodes the gravitational degrees
of freedom is a scalar field and, hence, its relation with the
four dimensional metric --specially in the quantum case-- is
somewhat indirect. Even though we have discussed some specific
issues of the combined system (such as two-point functions
involving both the scalar and gravitational field) we have mostly
used the matter field in this role of spacetime probe.

In order to obtain geometric information on quantum spacetime one
must try to work with objects of direct physical interpretation in
spacetime terms. One such object is the two-point function, in
fact, the field commutator has already been successfully used in
previous work of the authors to understand microcausality
\cite{BarberoG.:2003ye}. Here we have also considered two-point
functions with an approximate interpretation (as in ordinary
Minkowski space QFT) of propagation amplitudes from one spacetime
event to another. As is well known this interpretation is only
approximate because the Hilbert space vectors obtained by acting
on the vacuum state with the field operator do not form an
orthonormal set. In order to overcome this difficulty we have
introduced an orthonormal basis of position state vectors,
labelled by the radial coordinate, and used them to define radial
wave functions. These states are a generalization of the
Newton-Wigner states of ordinary QFT. With the aid of this
orthonormal basis we have considered first the Newton-Wigner
propagator $\langle R_2|\hat{U}(t_2,t_1)|R_1\rangle$ and we have
used it later to study the time evolution of a radial position
wave function.

The results obtained in all the approaches are different but
compatible. They can be summarized as follows:

\begin{itemize}

\item There are some interesting physical effects happening at the
symmetry axis. In particular all the approaches that we have
followed (the approximate ones provided by the two-point
functions, the ones given by the Newton-Wigner propagator, and the
radial wave functions themselves) suggest a significant
enhancement of the probability to find field quanta there. In
principle one could expect such a behavior due to backscattering
--as it happens when one considers the classical system--. However
we think that the comparison with the massless axially symmetric
field propagating in a Minkowskian background suggests that most
of it is due to a combination of quantum and gravitational
effects.

\item The probability amplitudes  show an enhanced probability of
finding scalar field quanta on some lines of slope 1 in the
$(\rho,\tau)$ plane that can be interpreted as approximate null
geodesics of an emergent metric. Even though we cannot obtain
other types of geodesics with our massless fields it is reassuring
to see the emergence of approximate spacetime trajectories with a
clear physical interpretation. This approximate null geodesics
correspond in this case to the ones of the Minkowski metric in 1+1
(or rather 2+1 with axial symmetry). It is worthwhile to note that
as the one particle states that we are using are ``the closest
ones" to the vacuum state (other than the vacuum state itself) we
are seing in an operational way the appearance of a classical flat
spacetime as far as (some) of its geometric properties are
concerned.

\item Another interesting effect is the persistence of the
amplitudes in the support of the initial wave function. This is
clearly displayed in the behavior of the wave function itself as
it evolves in time and shows up in the Newton-Wigner propagator. A
somewhat similar effect appears in the two-point function. The
divergence at $\rho_1=\rho_2$ for these objects is, however, a
consequence of the fact that we are not regularizing the fields.
Even if we regularize, for example by introducing a cut-off
\cite{BarberoG.:2004uv}, we would observe a large value for this
function at $\rho_1=\rho_2$.

\end{itemize}

It would be interesting to apply the methods developed in the
paper and use matter field quanta to explore quantized geometries
for states representing classical configurations corresponding to
arbitrary solutions for the Einstein-Rosen waves. In particular it
would be illuminating to compare the results with those obtained
by quantizing the matter fields in the curved backgrounds provided
by such solutions (where particle creation effects may play a
relevant role). This would first require us to find suitable
semiclassical states for the system describing a non trivial
gravitational part and a simple matter part and study their
quantum evolution. As long as they can be found our approach
should lead to unambiguous answers to questions related to the
emergence of classical trajectories for quantum tests particles.
We are working on this problem in the present moment. It should be
emphasized, however, that a direct comparison between both
approaches may be difficult because of the very different Hilbert
spaces used in their quantization. In particular the use of
Newton-Wigner states in the present scheme is very useful because
$n$-particle subspaces of our Hilbert space are stable under the
quantum evolution of the system. This allows us to rely on a
probabilistic interpretation of one-particle wave functions. For a
scalar field evolving in a general Einstein-Rosen curved
background this may be no longer possible due to particle creation
effects.

In our opinion the model provided by Einstein-Rosen waves, free or
coupled to cylindrically symmetric matter, is an excellent test
bed to discuss issues in quantum gravity. Of course there is
always the issue as to what extent the results obtained are
equivalent (at least in a qualitative way) to real effects in a
full theory of quantum gravity or artifacts of the symmetry
reduction. In this respect we think that the effects that we have
described above admit a sensible interpretation and give
interesting hints about the behavior of quantized gravity that we
want to explore in the future.

\begin{acknowledgments}
The authors want to thank Jos\'e Mar\'{\i}a Mart\'{\i}n
Garc\'{\i}a and Daniel G\'omez Vergel for their insightful
comments. I. Garay is supported by a Spanish Ministry of Science
and Education under the FPU program. This work is also supported
by the Spanish MEC under the research grant FIS2005-05736-C03-02.
\end{acknowledgments}

\begin{appendix}

\section{Numerical Computation of integrals involving two bessel
functions}{\label{app}}

Throughout the paper we have repeatedly encountered a class of
improper integrals involving products of Bessel functions. The
oscillatory character of the integrand, and the fact that they are
defined on the half-real line and are usually conditionally
convergent makes it necessary to find an efficient way to compute
them. We provide such a method here. This has been extensively
used in the several plots that appear in the paper. The seamless
mesh between these numerical results and the analytic
approximation provided by the asymptotic expansions is an eloquent
proof of both the accuracy of the asymptotic approximations and
the numerical results.

Let us consider integrals of the type
$$
I(\rho,\sigma,\tau):=\int_0^{\infty}J_{\mu}(\rho q)J_{\nu}(\sigma
q)f(q)\exp[-i\tau(1-e^{-q})]\,\mathrm{d}q
$$
with $\mu,\nu\in{\mathbb{Z}}$, and $f(q)$ a sufficiently regular
function such that the integral is, at least, conditionally
convergent. We first change variables according to $s=e^{-q}$ to
get
$$
I(\rho,\sigma,\tau)=e^{-i\tau}\int_0^{1}\frac{e^{i\tau
s}}{s}J_{\mu}(-\rho \log s)J_{\nu}(-\sigma \log s)f(-\log
s)\,\mathrm{d}s.
$$
We now  write it as the sum of two contour integrals, $I_1$ and
$I_2,$ in the complex $s$-plane defined on the paths
$C_1=\{iu:u\in[0,\infty)\}$, $C_2=\{1+iu:u\in[0,\infty)\}$:
\begin{eqnarray*}
&&I_1(\rho,\sigma,\tau)=e^{-i\tau}\int_0^{\infty}\frac{e^{-\tau
u}}{u}J_{\mu}(-\rho\log i u)J_{\nu}(-\sigma\log i u)f(-\log i u)\,\mathrm{d}u\,,\\
&&I_2(\rho,\sigma,\tau)=-i\int_0^{\infty}\frac{e^{-\tau u}}{1+i
u}J_{\mu}(-\rho\log [1+i u])J_{\nu}(-\sigma\log  [1+i u])f(-\log
 [1+i u])\,\mathrm{d}u\,.
\end{eqnarray*}
For the functions $f$ that appear in the paper the second integral
$I_2$ is very well behaved, because of the exponential fall-off of
the integrand and its non-singular character. It can be computed
numerically without difficulty. On the other hand the integrand in
$I_1$ has a nasty oscillating behavior in the vicinity of $u=0$
although the integral itself is convergent. A way to turn it into
a much tamer object is to use the contour integral representation
for the Bessel functions introduced above to write it as the
multiple integral
\begin{equation}
I_1(\rho,\sigma,\tau)=e^{-i\tau}\int_0^{\infty}\mathrm{d}u
\oint_{\gamma_1}\frac{\mathrm{d}z_1}{z_1^{\mu+1}}
\oint_{\gamma_2}\frac{\mathrm{d}z_2}{z_2^{\nu+1}}
\frac{f(-\log(iu))}{u}\,\exp\left(-\left[\frac{\rho}{2}
\left(z_1-\frac{1}{z_1}\right)+\frac{\sigma}{2}
\left(z_2-\frac{1}{z_2}\right)\right]\log(iu)-\tau u\right)
\label{multintegral}
\end{equation}
where $\gamma_{1,2}$ are simple closed paths surrounding the
origin in $\mathbb{C}$. By choosing these paths appropriately
(satisfying, for example, the conditions $\Re e(z_1-1/z_1)\leq 0$
and $\Re e(z_2-1/z_2)\leq 0$) it is possible in many cases to
guarantee that
$$
\int_0^{\infty}
\frac{f(-\log(iu))}{u}\,\exp\left(-\left[\frac{\rho}{2}
\left(z_1-\frac{1}{z_1}\right)+\frac{\sigma}{2}
\left(z_2-\frac{1}{z_2}\right)\right]\log(iu)-\tau
u\right)\,\mathrm{d}u
$$
is convergent. This allows us to change the integration order in
(\ref{multintegral}). Furthermore, for specific choices of the
function $f$ (most of the cases appearing in the paper!) this last
integral in $u$ can be exactly obtained in analytic form thus
leaving us with a double integral representation for $I_1$,
defined on a set with the topology of a torus, of a perfectly
regular function (except for a measure zero set of values of the
parameters $\rho$, $\sigma$, and $\tau$). For example if $\mu=0$,
$\nu=1$, $f=1$ we have
$$
I_1(\rho,\sigma,\tau)=\frac{e^{-i\tau}}{4\pi^2}
\oint_{\gamma_1}\!\frac{\mathrm{d}z_1}{z_1}\oint_{\gamma_2}\!\frac{\mathrm{d}z_2}{z_2^2}
\exp\left(\left[\frac{\rho}{2}\left(z_1-\frac{1}{z_1}\right)\!+\!\frac{\sigma}{2}
\left(z_2-\frac{1}{z_2}\right)\right](i\frac{\pi}{2}-\log\tau)\right)
\Gamma\left[\frac{\rho}{2}\left(z_1-\frac{1}{z_1}\right)\!+\!\frac{\sigma}{2}
\left(z_2-\frac{1}{z_2}\right)\right].
$$
The integrand is now well behaved and the integral can be computed
numerically in an efficient and quick way by ordinary methods.

\end{appendix}


\end{document}